# The Influence of Rough Surface Thermal-Infrared Beaming on the Yarkovsky and YORP Effects

B. Rozitis[a] and S. F. Green[a]

[a]*Planetary and Space Sciences, Department of Physical Sciences, The Open University, Walton Hall, Milton Keynes, MK7 6AA, UK*

No. of Manuscript Pages: 49
No. of Figures: 21
No. of Tables: 4

Please direct editorial correspondence and proofs to:

Benjamin Rozitis
Planetary and Space Sciences
Department of Physical Sciences
The Open University
Walton Hall
Milton Keynes
Buckinghamshire
MK7 6AA
UK

Phone: +44 (0) 1908 655808

Email: b.rozitis@open.ac.uk

Email address of co-author: s.f.green@open.ac.uk

## ABSTRACT

It is now becoming widely accepted that photon recoil forces from the asymmetric reflection and thermal re-radiation of absorbed sunlight are, together with collisions and gravitational forces, primary mechanisms governing the dynamical and physical evolution of asteroids. The Yarkovsky effect causes orbital semi-major axis drift and the YORP effect causes changes in the rotation rate and pole orientation. We present an adaptation of the Advanced Thermophysical Model (*ATPM*) to simultaneously predict the Yarkovsky and YORP effects in the presence of thermal-infrared beaming caused by surface roughness, which has been neglected or dismissed in all previous models. Tests on Gaussian random sphere shaped asteroids, and on the real shapes of asteroids (1620) Geographos and (6489) Golevka, show that rough surface thermal-infrared beaming enhances the Yarkovsky orbital drift by typically tens of percent but it can be as much as a factor of two. The YORP rotational acceleration is on average dampened by up to a third typically but can be as much as one half. We find that the Yarkovsky orbital drift is only sensitive to the average degree, and not to the spatial distribution, of roughness across an asteroid surface. However, the YORP rotational acceleration is sensitive to the surface roughness spatial distribution, and can add significant uncertainties to the predictions for asteroids with relatively weak YORP effects. To accurately predict either effect the degree and spatial distribution of roughness across an asteroid surface must be known.

# 1. INTRODUCTION

## 1.1 The Yarkovsky and YORP Effects

The asymmetric reflection and thermal re-radiation of sunlight from an asteroid's surface imposes a net force and torque. The net force (Yarkovsky effect) causes the asteroid's orbit to drift and the net torque (Yarkovsky-O'Keefe-Radzievskii-Paddack or YORP effect) changes the asteroid's rotation period and the direction of its spin axis (Bottke et al. 2006). Fig. 1 shows a schematic of the Yarkovsky and YORP effects acting on an ideal spherical asteroid with non-zero surface thermal inertia, rotating in a prograde sense, and with two wedges attached at different angles. Due to the surface thermal inertia the hotter surface temperatures are shifted away from the subsolar point leading to excess thermal emission on the afternoon side. This results in a net photon force pushing in the same direction as the orbital motion and thus causes the orbit to expand (Yarkovsky effect). If the asteroid were rotating in a retrograde sense then the situation would be reversed and the orbit would shrink. Since the magnitude of this effect is dependent on the asteroid rotation it is also referred to as the diurnal Yarkovsky effect. A seasonal Yarkovsky effect also exists for asteroids with non-zero obliquities, which is caused by the delayed thermal emission from the alternate heating of the two asteroid hemispheres. This form of the effect always causes the asteroid orbit to shrink, and it only becomes important for asteroids with very high thermal inertias. As also shown in Fig. 1, reflected sunlight and thermally emitted radiation from the two wedges creates photon torques acting in opposite directions about the asteroid centre of mass. However, as the wedges are mounted at different angles the reflection and emission directions are also different and the torques don't cancel out. The resultant torque increases or decreases the rotation rate, depending on the shape asymmetry, and can also shift the orientation of the spin axis (YORP effect). These combined effects are of fundamental importance for the dynamical and physical evolution of small asteroids in the solar system.

The Yarkovsky effect helps to deliver <40 km asteroids in the main belt to resonance zones capable of transporting them to Earth-crossing orbits, and dispersing asteroid families. It also adds significant uncertainties to predictions of the orbits of potentially hazardous asteroids during very close encounters with the Earth, such as (54509) Apophis (Giorgini et al. 2008). Currently, it has been detected by sensitive radar ranging for (6489) Golevka (Chesley et al. 2003), by deviations from predicted ephemerides over a long time span for (152563) 1992 BF (Vokrouhlický, Chesley & Matson 2008), and indirectly through the observed orbital distribution of the Karin cluster asteroid family (Nesvorný & Bottke 2004).

The YORP effect helps to explain the observed distribution of rotation rates in the asteroid population (Bottke et al. 2006); those smaller than 40 km, which are more susceptible to YORP-induced spin rate changes, have a clear excess of very fast and very slow rotators. YORP also causes the direction of the rotation axis to shift, although so far only indirect evidence for this effect exists through the clustering of rotation axis directions in asteroid families (Vokrouhlický, Nesvorný & Bottke 2003). Small gravitationally bound aggregates (rubble pile asteroids) could be spun up so fast that they are forced to change shape and/or undergo mass shedding (Holsapple 2010). Approximately 15% of near-Earth asteroids are inferred to be binaries (Pravec & Harris 2007), and YORP spin up is proposed as a viable formation mechanism (Walsh, Richardson & Michel 2008). In particular, the binary (66391) 1999 KW4 (Ostro et al. 2006) exhibits the typical physical and orbital characteristics (i.e. spinning-top primary shape and a spheroidal secondary with a near-circular equatorial orbit) predicted by YORP spin up. It has also been recently suggested that YORP spin up and fission of contact-binary asteroids is a viable formation mechanism of asteroid pairs (Pravec et al. 2010). The YORP effect has been detected for asteroids (54509)

YORP (Lowry et al. 2007; Taylor et al. 2007), (1862) Apollo (Kaasaleinen et al. 2007; Ďurech et al. 2008a), (1620) Geographos (Ďurech et al. 2008b), and (3103) Eger (Ďurech et al. 2009) by observing very small phase shifts in their rotational light curves over several years.

To accurately model the Yarkovsky and/or the YORP effect acting on an asteroid, any model must take into account the asteroid's size and shape, mass and moment of inertia, surface thermal properties, rotation state, and its orbit about the Sun. Past and current models tend to focus on modelling just one of the effects and not both simultaneously (see Table 1 for a list of models produced and used over the past decade). However, since the two effects are interdependent (as they each influence some of their dependent properties), a unified model of both Yarkovsky and YORP effects is required to accurately predict the long-term dynamical evolution of asteroids affected by them. For the selection of YORP-detected asteroids, the models generally overestimate the YORP effect acting on them when compared to the observations. Also, studies have shown the YORP effect to be highly sensitive to small-scale variations in an asteroid's shape model, and suggest that the error in any YORP effect prediction could have unity order (e.g. Breiter et al. 2009; Statler 2009). It remains uncertain whether these findings are a product of specific model assumptions or model simplifications.

Golevka was the first asteroid for which a Yarkovsky orbital drift has been detected and its semi-major axis has been measured to be decreasing at a rate of -95.6 ± 6.6 m yr$^{-1}$ (Chesley et al. 2003, 2008). It has been extensively studied using radar observations from which its orbital properties, shape, size, and rotation state have been derived (Hudson et al. 2000). By considering a range of typical asteroid surface thermal properties, Chesley et al. (2003) compared Yarkovsky model predictions with the observed orbital drift to infer that Golevka has a bulk density of 2700 $^{+400}/_{-600}$ kg m$^{-3}$. Numerous other YORP models have used Golevka's radar-derived shape for testing purposes but so far the YORP effect has not been measured (e.g. Čapek & Vokrouhlický 2004).

The Yarkovsky orbital drift on asteroid 1992 BF was discovered by a mismatch between precovery optical astrometric observations conducted in 1953 and the orbit predicted by observations spanning from 1992 to 2005 (Vokrouhlický, Chesley & Matson 2008). This mismatch couldn't be explained by inaccuracy of the 1953 observations, but it was found (using a simple linear heat diffusion model) that the Yarkovsky effect had the necessary strength to cause the observed orbital drift. 1992 BF's semi-major axis has been measured to be decreasing at a rate of -160 ± 10 m yr$^{-1}$.

Asteroid YORP (formerly known as 2000 PH5) was the first asteroid for which a YORP effect has been detected, which is causing its rotation to accelerate at a rate of 0.47 ± 0.05 rad yr$^{-2}$ (Lowry et al. 2007). Asteroid YORP already has a fast rotation period of 12.17 minutes but due to its YORP rotational acceleration it might reach a rotation period of just ~20 seconds towards the end of its expected dynamical lifetime. Taylor et al. (2007) conducted radar observations of asteroid YORP, deriving its irregular shape and rotation pole orientation allowing models to produce a theoretical rotational acceleration for comparison purposes. It was found that the model predictions overestimated the rotational acceleration compared with that observed by factors ranging from ~2 to ~6.

Apollo's YORP effect was detected shortly after that of asteroid YORP. Kaasaleinen et al. (2007) first measured its rotational acceleration, which Ďurech et al. (2008a) improved upon. Its rotation has been measured to be accelerating at a rate of (7.3 ± 1.6) x10$^{-3}$ rad yr$^{-2}$. This is consistent in sign and magnitude with the theoretical YORP value of 10.6 x10$^{-3}$ rad yr$^{-2}$, which was made using the light-curve-derived shape model and rotation pole orientation.

Geographos has a rotation that has been measured to be accelerating at a rate of (1.53 ± 0.20) x10$^{-3}$ rad yr$^{-2}$ (Ďurech et al. 2008b). Its shape has been determined by both radar (Hudson & Ostro 1999) and light curve (Ďurech et al. 2008b) inversion. The best fit light-curve-derived shape model predicts a YORP rotational acceleration of 1.87 x10$^{-3}$ rad yr$^{-2}$ i.e. the same sign and magnitude as that observed. Different variants of the light curve shape model predict YORP rotational accelerations ranging from 0.40 to 2.00 x10$^{-3}$ rad yr$^{-2}$ which straddle the observed value. However, the radar-derived shape model predicts a YORP rotational acceleration of -4.00 x10$^{-3}$ rad yr$^{-2}$ i.e. with a sign opposite to the observed value. Presumably, this is because the radar shape model is not unique as the data set it was derived from contains north-south ambiguities due to the near-equatorial view of the asteroid during the radar observations.

Eger is the latest asteroid to have a YORP effect detection and its rotation has been measured to be accelerating at a rate of (1.2 ± 0.8) x10$^{-3}$ rad yr$^{-2}$ (Ďurech et al. 2009). The shape model and rotation pole orientation derived from light curve inversion predicts a theoretical YORP value that is in agreement with the observations.

## 1.2 Overview of Previous Models

Generally, Yarkovsky/YORP (or YY for short) models fall into three categories: analytical, numerical, and semi-analytical. Analytical models represent the effects as mathematical closed-form solutions whereas numerical models use a time-stepping procedure to determine the YY effects' behaviour over time. Semi-analytical models combine elements of a numerical model with elements of an analytical one. Analytical and semi-analytical models are useful in that they are very quick to run and the functional dependence of certain parameters on the YY effects is obvious. However, they are generally first-order approximations and lack the higher accuracy of numerical model solutions. The problem with numerical models is that they can take a long time to run and the functional dependence of certain parameters can remain elusive. The decision on which model to use depends on the situation. If there is a lack of object information and/or a large number of runs is required for a statistical analysis then an analytical model may be the better option. However, if there is a lot of high quality information available about an object and/or an accurate prediction is required then a numerical model would be the best option.

In all types of model, the photon recoil forces and torques from absorbed, reflected, and thermally emitted photons are calculated by using the surface normals of the asteroid shape model. However, the way shape is represented differs between the models. Analytical models utilise either exact spheres or an expansion of spherical harmonics, whilst numerical models utilise polyhedrons made out of a number of triangular facets. For spheres the surface normal vector corresponds to the radial vector, for a spherical harmonic expansion the normal vectors are found by taking its gradient; and for a polyhedron the normal vectors are found perpendicular to the planes of the triangular facets. Shapes represented by spherical harmonic expansions are limited to deformed spheres and cannot represent very irregular shapes, in particular, where a radial vector can cross the surface more than once. Polyhedral representations do not have this limitation and any shape can be represented.

There is a lot of variation between the models in the method used to determine the surface temperature distribution, which is required for calculation of thermal emission photon recoil forces and torques. The simplistic approach is to assume zero thermal inertia so that the surface is in instantaneous equilibrium between its thermal emission and absorbed solar radiation. This is a suitable approximation, known as "Rubincam's Approximation", for calculating the YORP rotational acceleration acting on an asteroid, as previous models have shown it to be independent of thermal inertia (e.g. Čapek & Vokrouhlický 2004; Breiter, Bartczak & Czekaj 2010). However, it is not suitable for calculating the Yarkovsky effect and the YORP-induced rotation pole orientation shift as both are dependent on thermal inertia. It is also not suitable for calculations on meteoroids as heat can be conducted all the way across the body. At the other extreme, one YORP model has also considered infinitely-high thermal inertia such that the asteroid surface-temperature distribution becomes isothermal in longitude (Steinberg & Sari 2011). These two thermal inertia extremes bound the possible behaviours of the YORP-induced rotation pole orientation shift caused by thermal emission.

The effects of thermal inertia have been included by modelling heat diffusion in different ways. For asteroids much larger than the thermal skin depth lateral heat conduction can be ignored so that it is perpendicular to the surface and the 1D heat conduction equation applies. This is referred to in some models as the "plane parallel" case (e.g. Breiter, Bartczak & Czekaj 2010). For small asteroids and meteoroids comparable to the size of the thermal skin depth some models have included 3D heat conduction (e.g. Spitale & Greenberg 2001; Breiter, Vokrouhlický & Nesvorný 2010; Sekiya, Shimoda & Wakita 2012). In numerical models the 1D and 3D heat conduction equations are solved numerically using a finite difference method, whilst analytical models linearise the equations so that a first order mathematical function approximation can be obtained. In some YORP models the effects of thermal inertia are simply modelled as a time delay between absorption of solar radiation and emitted thermal radiation (e.g. Scheeres 2007; Scheeres & Mirrahimi 2008; Mysen 2008a,b). Two models consider varying thermal properties with depth (Breiter, Bartczak & Czekaj 2010; Čapek 2007), and only one considers temperature-dependent thermal properties (Čapek 2007).

In all types of model, the effects of horizon shadows (i.e. when the Sun is below the local horizon) are included. Numerical models treating non-spherical asteroids generally include the effects of projected shadows implemented by means of ray-triangle intersection tests i.e. a sunlight ray imposed on a particular triangular facet is checked to see whether it intersects another triangular facet before it reaches the facet of interest. This is a standard tool in computational geometry but is also computationally expensive and the number of calls to it must be minimised. Obviously there is no need to perform a test if the Sun is below the facet's horizon. Generally, most numerical models perform the tests once and store them in a lookup table but the way this is implemented differs between the models. Analytical models don't include the effects of projected shadows; however, it is possible for semi-analytical models to use numerically derived shadow maps.

## 1.3 Rough Surface Thermal-Infrared Beaming

None of the models include the effects of thermal-infrared beaming. Most models, as stated above, assume that the overall thermal emission from a surface is perpendicular to it and parallel with the surface normal. The Statler (2009) model did include a correction to the emission vector for surface elements whose sky is partly obscured by other parts of the asteroid surface. Since emission towards these other parts of the surface will be absorbed there is no net recoil force in these directions and so they do not contribute to the YY effects. The model also uses a hemispheric albedo derived from the optical scattering model of Hapke (2002) rather than the Bond albedo. However, for a surface with clear sky the overall thermal emission is still assumed to be perpendicular to the surface. Recent work by Breiter & Vokrouhlický (2011) has included full Hapke scattering theory in a numerical YORP model in order to investigate the influence of directional thermal emission caused by the regolith grains. Here it is referred to as microscopic beaming since it occurs on spatial scales (<1 mm) that are much smaller than the diurnal thermal skin depth (~1 cm) associated with the macroscopic roughness that causes the thermal-infrared beaming effect. Their results demonstrate that the YORP rotational acceleration predictions are almost insensitive to the scattering/emission model used but obviously the thermal-infrared beaming effect has been neglected due to the roughness scales considered.

Thermal-infrared beaming has the tendency to re-radiate absorbed solar radiation back towards the Sun in a non-lambertian way, and is caused by surface roughness occurring at scales ranging from the diurnal heat wave penetration depth (diurnal thermal skin depth) up to the spatial resolution of the detector used. It affects the observed asteroid thermal flux used in asteroid diameter determination, and has also been well observed in directional resolved thermal emission measurements of the lunar surface (Saari, Shorthill & Winter 1972) and in images taken of Comet 9P/Tempel 1 by Deep Impact (Groussin et al. 2007). The directional characteristics and causes of thermal-infrared beaming have been studied in detail by Rozitis & Green (2011) who produced a new rough surface thermophysical model called the Advanced Thermophysical Model (*ATPM*). The model was verified by accurately reproducing directional thermal emission measurements of the lunar surface and the inferred surface roughness was consistent with lunar roughness measured by other means. The study discovered that surface roughness and its associated thermal-infrared beaming effect moves the overall emission angle of thermal flux away from the surface normal. This obviously has implications for predicting the Yarkovsky and YORP effects as they are dependent on the directions of the photon recoil forces. Also, the study discovered that roughness alters the effective Bond albedo and thermal inertia of a surface, which has further implications for predicting the Yarkovsky effect as it is highly dependent on those properties.

Fig. 2 demonstrates how the rough surface thermal-infrared beaming effect works. For an ideal flat smooth surface heated by the Sun the resulting blackbody radiation exhibits Lambertian emission. The angular dependence of the smooth surface thermal emission follows a cosine law, as shown by the dashed line bubble, such that the net radiation vector is perpendicular to the surface. However, planetary surfaces are far from flat and exhibit surface roughness at all scales. If the surface is rough at scales similar to and greater than the thermal skin depth then the surface will produce a thermal-infrared beaming effect. As depicted here, those elements facing the Sun will be directly illuminated and heated (thick lines), whilst those facing away will be in shadow (thin lines). If each element is considered to be a Lambert emitter then thermal radiation will be directed back towards the Sun since the hottest elements are already facing it, as shown by the solid line bubble. Therefore, the net radiation vector is not always perpendicular to the surface as previous Yarkovsky and YORP models have assumed. Furthermore, as the rough surface elements are interfacing one another it leads to further exchange of heat via multiple scattering of sunlight and re-absorption of emitted thermal radiation. This increases the surface's capability of solar radiation absorption and heat retention which adds to the thermal-infrared beaming effect.

We have adapted the *ATPM* to simultaneously numerically model the Yarkovsky and YORP effects acting on an asteroid represented by a polyhedral shape model. It includes shadowing, emission vector corrections, thermal-infrared beaming, and global-selfheating. In this paper, we investigate the effects of thermal-infrared beaming due to macroscopic surface roughness. The effects of microscopic beaming are not included because experimental studies of directional thermal emission from Earth-based lava flows and sand surfaces show the thermal-infrared beaming effect to be predominantly caused by macroscopic surface roughness (Jakosky, Finiol & Henderson 1990), and that the YORP effect is seemingly insensitive to it (Breiter & Vokrouhlický 2011). Since surface temperatures are calculated by numerically solving the 1D heat conduction equation it is applicable only to asteroids much larger than the thermal skin depth. Table 1 compares the *ATPM* model features with other YY models developed and used over the past decade. The effects on the YY predictions when including rough surface thermal-infrared beaming effects for several shapes representative of asteroids are investigated and their implications discussed.

## 2. YARKOVSKY AND YORP MODELLING

### 2.1 Model Overview

The Yarkovsky and YORP effect model presented here is adapted from the *ATPM* thermophysical model which was developed to calculate the rough surface temperature distributions and thermal emissions of atmosphereless planetary bodies (Rozitis & Green 2011). Fig. 3 displays a schematic giving an overview of the physics and geometry used in the *ATPM*. The model accepts shape models in the triangular facet formalism to represent the global shape of a planetary body. It also accepts a topography model which it uses to represent the unresolved surface roughness in the global shape model for each facet. Any representation of the surface roughness can be used in the topography map but hemispherical craters are preferred since they have been shown to accurately reproduce the lunar thermal-infrared beaming effect and are easy to parameterise.

When evaluating the asteroid diurnal temperature variation, both types of facet (shape and roughness) are larger than the diurnal thermal skin depth (~1 cm) so that lateral heat conduction can be neglected and only 1D heat conduction perpendicular and into the surface can be considered. However, when evaluating the asteroid seasonal temperature variation only the shape facets are considered to be larger than the seasonal thermal skin depth (~1 to 10 m) for the 1D heat conduction approximation to still apply. Therefore, seasonal temperature variations are only evaluated for shape facets, and roughness facets are assumed to follow the same seasonal variations as their parent shape facets. The 1D heat conduction equation is solved with a surface boundary condition throughout: an asteroid rotation when evaluating the diurnal temperature variation for both types of facet, and an asteroid orbit when evaluating the seasonal temperature variation for only the shape facets.

For shape facets, the effects of selfheating from interfacing shape facets is neglected in this paper, and their surface boundary conditions just include direct solar radiation and shadowing. The effects of global-selfheating will be addressed in a future paper. For roughness facets, the effects of local-selfheating are included, and their surface boundary conditions include direct and multiple scattered solar radiation, shadowing, and re-absorbed thermal radiation from interfacing roughness facets. Shadowing of both types of facet is determined by standard ray-triangle intersection tests, and the radiative heat transfer problem of heat exchange between interfacing roughness facets is solved by using view factors.

Depending on the assumed surface properties the model requires between 10 and 100 revolutions to converge to a solution. Once the facet illumination fluxes and temperature variations are calculated they are transformed to reflected and thermally emitted photon recoil forces. These forces are directed along vectors that are anti-parallel to the surface normal of facets, or along vectors that take into account the surface normal of facets and directions where reflected/emitted photons are re-absorbed. The net force for any shape facet will therefore not generally be anti-parallel to its surface normal because of surface roughness. Photon torques are then calculated by taking the cross product of the photon force vectors with the facet position vectors about the asteroid centre of mass.

The degree of surface roughness for each shape facet is specified by a roughness fraction, $f_R$, that dictates the fraction of the shape facet represented by the rough surface topography model, and the remaining fraction, $(1 - f_R)$, represented by a smooth and flat surface. Since each shape facet has an individually assigned roughness fraction it enables different surface roughness distributions to be created across a modelled asteroid surface. The total photon force/torque for each shape facet is calculated by weighting the rough and smooth components by these fractions, and then the total photon force/torque for the asteroid is calculated by summing over all shape facets. The total asteroid photon force/torque is transformed into Yarkovsky and YORP accelerations by averaging over the asteroid rotation and orbit, and by taking into account the asteroid mass, moment of inertia, and pole orientation. Certain aspects of this modelling process are described in more detail in the following subsections.

## 2.2 Thermal Modelling

The temperature $T$ for each shape and roughness facet is determined by solving the energy balance equation, which leads to the surface boundary condition

$$(1 - A_B)\left([1 - S(\tau)]\psi(\tau)F_{SUN} + F_{SCAT}(\tau)\right) + (1 - A_{TH})F_{RAD}(\tau) + \frac{\Gamma}{\sqrt{4\pi P}}\left(\frac{dT}{dz}\right)_{z=0} - \varepsilon\sigma T_{z=0}^4 = 0, \quad (1)$$

where $\varepsilon$ is the emissivity, $\sigma$ is the Stefan-Boltzmann constant, $A_B$ is the Bond albedo, $S(\tau)$ indicates whether the facet is shadowed at normalized time $\tau$, $A_{TH}$ is the albedo at thermal-infrared wavelengths, $P$ is the rotational (diurnal-Yarkovsky effect) or orbital (seasonal-Yarkovsky effect) period, and $z$ is the normalized depth below the asteroid surface. $\Psi(\tau)$ is a function that returns the cosine of the Sun illumination angle at normalized time $\tau$, and $F_{SUN}$ is the integrated solar flux at the distance of the asteroid. The normalized time and depth, $\tau$ and $z$, are related to the actual time and depth, $t$ and $x$, via

$$\tau = \frac{t}{P} \qquad z = \frac{x}{l_d} \qquad (2)$$

where $l_d$ is the diurnal or seasonal thermal skin depth. The surface thermal inertia $\Gamma$ is related to the thermal conductivity $k_c$, specific heat capacity $C_h$, and density $\rho_s$ of the asteroid surface material via

$$\Gamma = \sqrt{k_c \rho_s C_h} \ . \qquad (3)$$

Interfacing facets will receive additional flux contributions from multiple-scattered sunlight and re-absorbed thermal emission from neighbouring facets. $F_{SCAT}(\tau)$ and $F_{RAD}(\tau)$ are then the total multiple-scattered and re-emitted thermal fluxes incident on a facet respectively at normalised time $\tau$. For shape facets where no selfheating occurs these last two terms are zero.

Heat conduction in the absence of an internal heat source can be described by the normalized 1D heat conduction (diffusion) equation

$$\frac{\partial T}{\partial \tau} = \frac{1}{4\pi}\frac{\partial^2 T}{\partial z^2} \ , \qquad (4)$$

and since the amplitude of sub-surface temperature variations decreases exponentially with depth, it implies an internal boundary condition given by

$$\left(\frac{\partial T}{\partial z}\right)_{z \to \infty} \to 0 \ . \qquad (5)$$

A finite difference numerical technique is used to solve the problem defined by equations 1, 4, and 5, and a Newton-Raphson iterative technique is used to solve the surface boundary condition (full details of which are given in Rozitis & Green (2011)). When evaluating the diurnal temperature variation at a single orbital point, the $[1 - S(\tau)]\Psi(\tau)$, $F_{SCAT}(\tau)$, and $F_{RAD}(\tau)$ terms of equation 1 are functions of rotational phase. However, when evaluating the seasonal temperature variation, they are functions of orbital phase and their values are determined by averaging their rotational phase counterparts at each orbital point. This is the same as the approach developed by Vokrouhlický & Farinella (1998) for numerically evaluating the seasonal Yarkovsky effect. Typically, the model uses 400 time steps and 40 depth steps going to a maximum depth of 1 thermal skin depth to solve the problem defined here.

The additional flux contributions from multiple-scattered sunlight and re-absorbed thermal emission are calculated using view factors. The view factor from facet $i$ to facet $j$, $f_{i,j}$, is defined as the fraction of the radiative energy leaving facet $i$ which is received by facet $j$ assuming Lambertian emission (Lagerros 1998). It is

$$f_{i,j} = v_{i,j} \frac{\cos\theta_i \cos\theta_j}{\pi d_{i,j}^2} a_j, \qquad (6)$$

where $v_{i,j}$ indicates whether there is line-of-sight visibility between the two facets, $\theta_i$ is facet $i$'s emission angle, $\theta_j$ is facet $j$'s incidence angle, $d_{i,j}$ is the distance separating facet $i$ and $j$, and $a_j$ is the surface area of facet $j$. However, the view factor given by equation 6 is an approximation since it applies to large separation distances relative to the facet area, and can become very inaccurate when the relative separation distances are very small. A simple method to calculate a more accurate view factor between any two facets in such cases is to split them up into a number of equal area subfacets, $MM$, and determine the view factors associated with each subfacet combination (introduced in Rozitis & Green (2011) as "better viewfactors"). The effective overall view factor is then

$$f_{i,j} = \frac{1}{a_i} \sum_{v=1}^{MM} \left( a_{iv} \sum_{u=1}^{MM} f_{iv,ju} \right), \qquad (7)$$

where $a_{iv}$ is the area of subfacet $iv$ which is a part of facet $i$, and $f_{iv,ju}$ is the view factor from subfacet $iv$ to subfacet $ju$ as calculated by equation 6.

Utilising the view factors, the multiple-scattered flux leaving facet $i$, $G_i(\tau)$, can be written as

$$G_i(\tau) = A_B \left( F_{SUN} \left[ 1 - S_i(\tau) \right] \psi_i(\tau) + \sum_{j \neq i} f_{i,j} G_j(\tau) \right), \qquad (8)$$

which can be efficiently solved using a Gauss-Seidel iteration, and the multiple-scattered flux incident on a facet is then

$$F_{SCAT}(\tau) = \frac{G(\tau)}{A_B}. \qquad (9)$$

The total incident re-emitted thermal flux can be calculated using

$$F_{RAD}(\tau) = \varepsilon\sigma \sum_{j \neq i} f_{i,j} T_j^4(\tau), \qquad (10)$$

where $T_j(\tau)$ is the surface temperature of facet $j$ at normalized time $\tau$. Since planetary surfaces absorb most of the incoming radiation at thermal-infrared wavelengths, i.e. $A_{TH} \sim 0$, only single scattering is considered.

## 2.3 Thermal-Infrared Beaming Photon Forces and Torques

There are three types of photons that can impose a recoil force and torque on an asteroid surface: absorbed solar, reflected solar, and thermally radiated. The absorbed solar photon recoil force is simply solar radiation pressure. It acts along the Sun-asteroid vector, which results in an effectively smaller value of the Sun's gravitational mass, and leads to a very tiny modification of the asteroid orbital period. For typical asteroids, its orbital perturbation is negligible compared to that produced by thermally radiated photon recoil forces (Žižka & Vokrouhlický 2011). Previous work has also determined that absorbed solar photon torques cancel out over an asteroid orbit resulting in no net torque regardless of the asteroid shape (Nesvorný & Vokroulický 2008; Rubincam & Paddack 2010). Yarkovsky force and YORP torque calculations can therefore neglect absorbed solar photons for simplicity.

Both reflected solar and thermally radiated photons are assumed to have isotropic (Lambert) emission profiles from a smooth flat facet. The photon recoil force acting on facet $i$, $\boldsymbol{p}_i(\tau)$, is therefore given by

$$p_i(\tau) = -\frac{2E_i(\tau)a_i}{3c}n_i(\tau),$$ (11)

where $a_i$ and $n_i(\tau)$ are the facet area and normal respectively, and $c$ is the speed of light. $E_i(\tau)$ is the radiant emittance of the facet, which is $G_i(\tau)$ for reflected solar photons and $\varepsilon\sigma T_i^4(\tau)$ for thermally radiated photons. This equation is true only if the facet has no other facets visible to it above its local horizon. If other facets are visible then photons emitted towards these facets will be re-absorbed resulting in an absorption recoil force that cancels out its emittance recoil force. If $f_{i,j}(\tau)$ is the unit vector associated with view factor $f_{i,j}$ giving the direction from facet $i$ to facet $j$ then the re-absorbed photon recoil force $q_{i,j}(\tau)$ is given by

$$q_{i,j}(\tau) = \frac{E_i(\tau)a_i}{c}f_{i,j}f_{i,j}(\tau).$$ (12)

The total photon recoil force including re-absorbed photons is then

$$p_i(\tau) = -\frac{2E_i(\tau)a_i}{3c}n_i(\tau) + \sum_{j\neq i}q_{i,j}(\tau),$$ (13)

which combined with equation 12 can be rearranged to

$$p_i(\tau) = -\frac{2E_i(\tau)a_i}{3c}\left(n_i(\tau) - \frac{3}{2}\sum_{j\neq i}f_{i,j}f_{i,j}(\tau)\right).$$ (14)

If the view factor between two facets has been calculated by the better view factor method then the unit vector associated with that view factor in that case is given by

$$f_{i,j}(\tau) = \frac{1}{a_i}\sum_{v=1}^{MM}\left(a_{iv}\sum_{u=1}^{MM}f_{iv,ju}f_{iv,ju}(\tau)\right),$$ (15)

where $f_{iv,ju}(\tau)$ is the unit vector pointing from subfacet $iv$ to subfacet $ju$.

For shape facets the recoil forces given by equation 11 are calculated in the heliocentric ecliptic coordinate system for evaluation of both the Yarkovsky and YORP effects. A fixed-frame asteroid-centric coordinate system is more preferable for evaluation of just the YORP effect but the heliocentric ecliptic system is used to avoid interchanging between the two. However, for convenience when dealing with roughness facets it is easier to calculate the recoil forces in a coordinate system where their normal vectors do not change with time. A suitable system is where the x and y axes define a plane that lies parallel to the plane of a parent shape facet with the x axis lying parallel to the vector pointing from the zeroth-vertex to the first-vertex of the shape facet. The z axis is therefore perpendicular to this plane and lies parallel to the shape facet normal (this coordinate system was defined as the "surface-roughness coordinate system" in Rozitis & Green (2011)). The summed roughness recoil forces in the surface-roughness coordinate system for shape facet $k$, $p_{rough,SR,k}(\tau)$, is then

$$p_{rough,SR,k} = \sum_{l=1}^{M}p_{SR,k,l}(\tau),$$ (16)

where $p_{SR,k,l}(\tau)$ are the recoil forces for roughness facet $l$ (for $l = 1$ to $M$ roughness facets) in the surface-roughness coordinate system (which is denoted by $SR$). The summed roughness recoil forces can be transformed into the heliocentric ecliptic coordinate system (which is denoted by $HE$), $p_{rough,HE,k}(\tau)$, by

$$p_{rough,HE,k}(\tau) = ACF_k\big(M_k(\tau)\cdot p_{rough,SR,k}(\tau)\big),$$ (17)

where $M_k(\tau)$ is a transformation matrix between the surface-roughness and heliocentric ecliptic coordinate systems (specified in Appendix A of Rozitis & Green (2011)). The $ACF_k$ term is an area conversion factor which ensures that the total area of the rough surface topography model projected along the shape facet normal has the same area as the shape facet (the topography model may not necessarily be defined in the same units as the global shape model).

The smooth and rough surface recoil forces for shape facet $k$ can be combined as a function of its roughness fraction $f_{R,k}$ to give the total recoil force, $p_{total,HE,k}(\tau)$, as

$$p_{total,HE,k}(\tau) = (1 - f_{R,k})p_{smooth,HE,k}(\tau) + f_{R,k}\,p_{rough,HE,k}(\tau), \qquad (18)$$

where the smooth component, $p_{smooth,HE,k}(\tau)$, has been calculated by equation 11. Finally, the photon torque associated with the total recoil force for shape facet $k$, $\varphi_{total,HE,k}(\tau)$, is then

$$\varphi_{total,HE,k}(\tau) = r_k(\tau) \times p_{total,HE,k}(\tau), \qquad (19)$$

where $r_k(\tau)$ is shape facet $k$'s position vector about the asteroid centre of mass.

## 2.4 Evaluation of the Yarkovsky and YORP Effects

The total photon force and torque acting on an asteroid at a specific point $i$ in its orbit, $P_i$ and $\Phi_i$, can be calculated by summing the recoil forces and torques from each shape facet across the asteroid surface and then by rotation-averaging. For determination of the Yarkovsky orbital drift the total force can be split into components that act along: the Sun-asteroid vector, the vector defining the plane of the orbit, and the vector perpendicular to these two. These force vectors have magnitudes $P_{x,i}$, $P_{z,i}$, and $P_{y,i}$ respectively. For a circular orbit and an orbit-perpendicular asteroid pole orientation, the rate of change in semi-major axis, $da/dt$, caused by the Yarkovsky effect can be evaluated by using just one of the latter quantities in

$$\frac{da}{dt} = \frac{2P_y}{\eta M_{AST}}, \qquad (20)$$

where $\eta$ is the orbital mean motion and $M_{AST}$ is the asteroid mass. This is applicable since the $P_y$ force component does not change throughout an orbit in this situation. However, for an asteroid on an elliptical orbit and/or with a pole orientation not perpendicular to the orbital plane this can vary throughout an orbit, which requires additional orbit-averaging for accurate evaluation of the Yarkovsky orbital drift. The orbit-average rate of change in semi-major axis for a general orbit is given by

$$\frac{da}{dt} = \frac{2a^2}{P_{ORB}GM_{SUN}M_{AST}}\Delta E, \qquad (21)$$

where $a$ and $P_{ORB}$ are the semi-major axis and period of the orbit respectively, $G$ is the gravitational constant, and $M_{SUN}$ is the mass of the Sun. $\Delta E$ is the total change in orbital energy integrated over one orbit, and can be found by the summation over $n$ orbital positions

$$\Delta E = \sum_{i=1}^{n} \Delta t_i \left( v_{x,i} P_{x,i} + v_{y,i} P_{y,i} + v_{z,i} P_{z,i} \right), \qquad (22)$$

where $\Delta t_i$ is the time spent at each orbital position, and $v_i$ are the orbital velocity components for the directions defined by the $P_i$ force components. As $v_z$ is always zero in this geometry (i.e. it is directed perpendicular to the orbital plane) then the $P_z$ component of the total photon force can be neglected from the calculation of orbital semi-major axis drift.

The YORP torques can also be transformed into suitable meaningful components. These are the rate of change in angular velocity (rotational acceleration), $d\omega_i/dt$, the rate of change in obliquity, $d\xi_i/dt$, and the precession in longitude, $d\lambda_i/dt$ (Bottke et al. 2006). Using the total torque, $\Phi_i$, these can be calculated by

$$\frac{d\omega_i}{dt} = \frac{\Phi_i \cdot d}{C_\omega} \qquad (23)$$

$$\frac{d\xi_i}{dt} = \frac{\Phi_i \cdot d_{\perp 1}}{C_\omega \omega} \qquad (24)$$

$$\frac{d\lambda_i}{dt} = \frac{\mathbf{\Phi}_i \cdot \mathbf{d_{\perp 2}}}{C_\omega \omega} \qquad (25)$$

where $C_\omega$ is the asteroid moment of inertia about its shortest axis, $\omega$ is the angular rotation rate, and $\mathbf{d}$ is the unit vector of the rotation pole direction. $\mathbf{d_{\perp 1}}$ and $\mathbf{d_{\perp 2}}$ are the unit vectors

$$\mathbf{d_{\perp 1}} = \frac{(\mathbf{o} \cdot \mathbf{d})\mathbf{d} - \mathbf{o}}{\sin \xi} \qquad (26)$$

$$\mathbf{d_{\perp 2}} = \frac{\mathbf{d} \times \mathbf{o}}{\sin \xi} \qquad (27)$$

where $\mathbf{o}$ is the unit vector defining the plane of the asteroid orbit. For calculation of rotational acceleration, equation 23 can be directly applied to an asteroid on a circular orbit with a pole orientation perpendicular to its orbital plane by evaluating the rotational torque component at one point in its orbit. This is applicable since the strength and sign of the rotational component does not change throughout an orbit. If the asteroid is on an elliptical orbit with a pole orientation still perpendicular to its orbital plane then the rotational component strength varies inversely proportional to the square of the heliocentric distance. In this case, the calculation of rotational acceleration for a circular orbit can be corrected to an elliptical orbit with the same semi-major axis by multiplying by the factor $YCF$ (Scheeres 2007),

$$YCF = \left(1 - e^2\right)^{-\frac{1}{2}}. \qquad (28)$$

However, for evaluation of the other components of YORP torque, and/or for evaluation of the rotational acceleration for an asteroid with a pole orientation not perpendicular to its orbital plane, then the varying YORP torque strength and sign throughout an orbit must be taken into account. In these cases, the YORP torque must be orbit-averaged over $n$ orbital positions using

$$\frac{dY}{dt} = \frac{1}{P_{ORB}} \sum_{i=1}^{n} \Delta t_i \frac{dY_i}{dt}, \qquad (29)$$

where $Y$ denotes the three different components of YORP torque, and $dY_i/dt$ is the YORP torque strength at orbital position $i$ given by equations 23 to 25.

## 2.5 Surface Roughness Distributions

The surface roughness that causes thermal-infrared beaming in each shape facet is represented by a hemispherical crater consisting of 100 roughness facets (see Fig. 4). It can accurately recreate the beaming effect produced by a range of surface roughness spatial scales simultaneously, and has been shown to reproduce the lunar beaming effect caused by surface roughness ranging from centimetre to 10-kilometre spatial scales (Rozitis & Green 2011). This is because it is almost size independent, and a rough shape facet can be considered to contain either multiple craters or a single crater with the same total cross-sectional area since they both produce the same effect. As each shape facet has an individual roughness fraction $f_R$ it is possible to create different surface roughness distributions across an asteroid surface. In this work, three different types of distributions are considered: 'uniform', 'normal', and 'patchy normal'. 'Uniform' distributions have the same roughness fraction assigned to each shape facet. This kind of distribution has always been assumed for thermophysical models used in interpreting thermal-infrared observations of asteroids (e.g. Spencer 1990; Müller & Lagerros 1998; Delbo' & Tanga 2009). However, the degree of surface roughness is likely to vary across an asteroid surface, although the degree of variation for any planetary body at scales similar to the thermal skin depth has never been measured. The 'normal' and 'patchy normal' distributions attempt to model varying surface roughness in a sensible way. 'Normal' distributions have independent random roughness fractions assigned to each shape facet chosen from a normal distribution with a specified mean and width. 'Patchy normal' distributions divide the asteroid surface into 10 randomly-chosen regions, and then each shape facet within a region is allocated the same roughness fraction which is also randomly-chosen from a normal distribution. Five different degrees of surface roughness are also considered: 'zero', 'low', 'medium', 'high', and 'full'. These correspond to mean roughness fractions of 0.00, 0.25, 0.50, 0.75, and 1.00 respectively. Finally, two different widths are considered for the normal distribution: 'narrow' and 'wide'. The 'narrow' and 'wide' distributions have $3\sigma$ widths that are 0.25 and 0.50 roughness fractions respectively. Fig. 5 and Table 2 summarise the properties of the different surface roughness distributions considered.

## 3. PARAMETER SENSITIVITY

### 3.1 Conventional Parameters

Before the effects of thermal-infrared beaming on the Yarkovsky/YORP predictions is studied, it is useful to first review their dependence on already known important parameters for comparison purposes. In this section the parameter dependence is investigated using the shape model of near Earth asteroid Geographos. The model assumes a circular orbit and the other parameters are given values equal to those observed by various means or are varied to study their model dependence. The results presented in this section are therefore not accurate predictions of the effects acting on Geographos and thereafter the model is referred to as "pseudo-Geographos".

For the pseudo-Geographos model, we adopt the light curve shape model of Ďurech et al. (2008b). The model is scaled so that it has a volume equal to a 2.56 km diameter sphere, which gives it a maximum equatorial diameter consistent with the value of $5.0 \pm 0.15$ km determined from radar observations (Hudson & Ostro 1999). A rotation period of 5.223 hours, determined from the light curve observations, and an orbital semi-major axis of 1.246 AU are also used. Assuming a bulk density of 2500 kg m$^{-3}$ gives a mass of 2.2 x10$^{13}$ kg and a moment of inertia about the short axis of 3.4 x10$^{19}$ kg m$^2$. Fig. 6 depicts the shape and a typical model surface temperature distribution of pseudo-Geographos, and Table 3 lists the properties used.

For a Bond albedo of 0.1 and a thermal inertia of 200 J m$^{-2}$ K$^{-1}$ s$^{-1/2}$ (i.e. the near Earth asteroid average thermal inertia determined by Delbo' et al. (2007)), pseudo-Geographos has a Yarkovsky semi-major axis drift of 18.1 m yr$^{-1}$ and a YORP rotational acceleration of 2.26 x10$^{-3}$ rad yr$^{-2}$ at 0° obliquity. The Yarkovsky force is proportional to the cross-sectional area and thus the square of the diameter, but the mass is proportional to the volume and thus the cube of the diameter. Therefore, the Yarkovsky semi-major axis drift is inversely proportional to the diameter $D$

$$\frac{da}{dt} \propto \frac{1}{\rho_b D}. \tag{30}$$

The YORP torque is proportional to the asteroid area and also to the radius and thus the cube of the diameter, but the moment of inertia is proportional to the fifth power of the diameter. Therefore, the YORP rotational acceleration is inversely proportional to the square of the diameter

$$\frac{d\omega}{dt} \propto \frac{1}{\rho_b D^2}. \tag{31}$$

For example, the small uncertainty of $\pm$ 3% on Geographos's radar-derived diameter introduces uncertainties of $\pm$ 3% and $\pm$ 6% on pseudo-Geographos's Yarkovsky and YORP predictions respectively. Both effects are also inversely proportional to the asteroid bulk density $\rho_b$.

The different surface temperature distributions caused by different values of thermal inertia and rotation are specified by the dimensionless thermal parameter, $\Theta$, given by

$$\Theta = \frac{\Gamma \sqrt{\omega}}{\varepsilon \sigma T_{SSP}^3}, \tag{32}$$

where $T_{SSP}$ is the subsolar point temperature in which zero thermal inertia and no rotation are assumed. Different values of thermal inertia and rotation period can combine to give the same thermal parameter value, and when they do they result in the same surface temperature distribution (Spencer, Lebofsky & Sykes 1989). By using this property, the rotation dependence of the YY predictions can be studied simultaneously when studying the thermal inertia dependence. Fig. 7a shows the Yarkovsky effect acting on pseudo-Geographos as a function of thermal inertia and Bond albedo at 0° obliquity. The thermal inertia has been chosen to range from 0 J m$^{-2}$ K$^{-1}$ s$^{-1/2}$ (i.e. instantaneous equilibrium) representing a fine dusty surface, to 2000 J m$^{-2}$ K$^{-1}$ s$^{-1/2}$ representing a surface consisting of exposed bare rock. The Bond albedo was given three different values of 0.1, 0.3, and 0.5 ranging from optically dark to very reflective asteroids. Fig. 7a indicates that the Yarkovsky effect is, as expected, highly dependent on the thermal inertia and therefore also the rotation period. There is no Yarkovsky effect at zero thermal inertia as there is no morning-afternoon surface temperature asymmetry. It also indicates that the Yarkovsky effect peaks at ~200 J m$^{-2}$ K$^{-1}$ s$^{-1/2}$ for pseudo-Geographos. After this peak the Yarkovsky effect decreases towards zero as thermal inertia increases. This is because increasing thermal inertia smoothes out the temperature distribution in longitude and thus reduces the morning-afternoon surface temperature asymmetry. Although not shown, the YORP rotational acceleration is independent of thermal inertia and therefore also the thermal parameter and rotation period. It maintains the same value across the range and any small deviations observed are caused by the numerical accuracy of the model rather than any physical relationship. Câpek & Vokroulický (2004) first explained this thermal inertia independence; since equation 23 indicates that the YORP rotational acceleration is determined by the torque projection on to the rotation axis, it basically depends on the total amount of energy thermally reprocessed at a given latitude of an asteroid. Thermal inertia affects the delay between when sunlight is absorbed and then thermally re-emitted, but not the total amount of this energy. Rotation-averaging then removes rotation-phase differences between predictions of different thermal inertias and explains this result.

As expected, the Yarkovsky effect decreases with increasing Bond albedo because as the Bond albedo increases, less incident solar flux is absorbed by the asteroid surface and emitted away as thermal radiation. However, the Yarkovsky effect is not quite directly proportional to $(1 - A_B)$ as one might expect because the Bond albedo also appears in the thermal parameter (within the subsolar point temperature term) and influences the surface temperature distribution. As shown in Fig. 7a the thermal inertia at which the Yarkovsky effect peaks decreases slightly with increasing Bond albedo. The YORP rotational acceleration which is known to be independent of thermal inertia is also independent of Bond albedo, at least in cases where the reflected solar flux and the thermal radiation have the same directional distribution. In this case it is assumed to be Lambertian, and the Bond albedo simply determines the fractions of the YORP torque arising from each component. For example, a Bond albedo of 0.1 will lead to 10% of the total torque caused by reflected solar flux and 90% caused by thermal radiation, and for a Bond albedo of 0.5 each component contributes 50% of the total torque.

Fig. 7b shows pseudo-Geographos's Yarkovsky effect at a number of heliocentric distances as a function of thermal inertia at 0° obliquity. Since the solar flux is inversely proportional to the square of the heliocentric distance, one might also expect the Yarkovsky effect to be inversely proportional to the square of the heliocentric distance. Fig. 7b shows that this is not quite the case; the Yarkovsky effect doesn't decrease as sharply for two reasons. The incident solar flux influences the subsolar point temperature and therefore the thermal parameter, which dictates the surface temperature distribution, and the Yarkovsky force produces a larger orbital perturbation at larger heliocentric distance. The Yarkovsky effect (as measured by semi-major axis drift) peaks at lower thermal inertia values with increasing heliocentric distance because of this. Also not shown, the YORP rotational acceleration at 0° obliquity is inversely proportional to the square of the heliocentric distance because it is directly proportional to the incident solar flux and is independent of the thermal parameter.

Fig. 7c shows pseudo-Geographos's diurnal and seasonal Yarkovsky effect as a function of obliquity. The near-Earth asteroid average thermal inertia of 200 J m$^{-2}$ K$^{-1}$ s$^{-1/2}$ (Delbo' et al. 2007) is assumed for the diurnal effect, and 2000 J m$^{-2}$ K$^{-1}$ s$^{-1/2}$ is assumed for the seasonal effect in order to maximise it. As the seasonal effect is quite negligible compared to the diurnal effect, it has been multiplied by a factor of -100 so that it can be plotted on the same axes as the diurnal effect. The minus sign is required to change it from a decrease to an increase in semi-major axis. As shown, the diurnal effect is maximised at zero obliquity, whilst the seasonal effect is maximised at 90° obliquity. Fig. 7d shows the YORP effect as a function of obliquity. There is a rotational acceleration at zero obliquity and a rotational deceleration at 90° obliquity, and there is a gradual transition between the two for the in between obliquities. There is a zero YORP rotational acceleration crossing point at ~55° obliquity, which is referred to as the critical angle. Many objects whose YORP effects that have been studied as a function of obliquity exhibit crossing points close to this value (e.g. Rubincam 2000; Vokrouhlický & Čapek & 2002; Micheli & Paolicchi 2008). As summarised by Micheli & Paolicchi (2008), Nesvorný & Vokrouhlický (2007) show analytically that it is caused by a complex coupling between insolation and longitudinal shape irregularities, and is also dependent on the spectrum of such irregularities.

In terms of the relative impact on the YY predictions, the asteroid size, bulk density, and obliquity have the largest influences. For the Yarkovsky effect, in order of importance, the heliocentric distance, the thermal inertia, the rotation period, and the Bond albedo have an interdependent influence on the prediction as they are all related to one another via the thermal parameter. The YORP rotational acceleration is not influenced by the thermal parameter and is thus independent of thermal inertia, rotation period, and Bond albedo. However, it is inversely proportional to the square of the heliocentric distance.

## 3.2 Rough Surface Thermal-Infrared Beaming

The influence of rough surface thermal-infrared beaming on the YY predictions is studied using different shape models that are representative of asteroid shapes. These include Gaussian-sphere shape models, and the shape models of asteroids Geographos and Golevka. Gaussian-sphere shape models have been extensively used in other works to study the parameter dependence of the YORP effect (e.g. Vokrouhlický & Čapek 2002; Čapek & Vokrouhlický 2004; Statler 2009), and Appendix A briefly outlines how to create them. The radar-derived shape model of Golevka and the light-curve-derived shape model of Geographos are used because they have more extreme shapes than the typical Gaussian-spheres and have measured Yarkovsky and YORP effects respectively (Chesley et al. 2003; Ďurech et al. 2008b). Fig. 6 displays shape models of two example Gaussian-spheres and those of asteroids Geographos and Golevka. The albedo, obliquity, and thermal inertia dependencies of the YY predictions, subject to the different surface roughness distributions, is studied. To characterise the degree of scatter in the predictions caused by surface roughness, predictions are made for 1000 independent realisations of each type of surface roughness distribution considered, which are then used to find the average prediction and its standard deviation. For simplicity, and like other works, a circular orbit is assumed to remove the dependence of the rotation pole longitude parameter.

### 3.2.1 Gaussian-Sphere Asteroids

Using the method outlined in Appendix A, ten different Gaussian-sphere shape models were created in addition to a standard sphere. Each shape model consists of 1152 facets which represents a good compromise between shape accuracy and computation time required, and is also consistent with the number used in other works (e.g. Vokrouhlický & Čapek 2002; Micheli & Paolicchi 2008). Each artificial asteroid is scaled so that it has a volume equal to a 1 km diameter spherical asteroid, and each given a bulk density of 2500 kg m$^{-3}$. They are also each given an orbital semi-major axis of 1 AU and a rotation period of 6 hours. Table 3 summarises the model parameters used.

Whilst evaluating the Yarkovsky effect for each Gaussian-sphere asteroid it was found that the predictions of semi-major axis drift only deviated by ~10% on average from the prediction for a sphere. This indicates that the Yarkovsky effect is not that sensitive to differences in shape for asteroids of the same size. Thus, presented in Fig. 8 and Fig. 9 are the 'shape-average' parameter dependencies of the Yarkovsky effect for the 11 different shaped asteroids (i.e. the mean of the results for the 11 different shape models with thermal and physical parameters as indicated). Fig. 8a shows the 'shape-average' Yarkovsky effect as a function of obliquity and 'uniform' surface roughness, at fixed thermal inertia and Bond albedo of 200 J m$^{-2}$ K$^{-1}$ s$^{-1/2}$ and 0.1 respectively. Fig. 8b shows the 'shape-average' Yarkovsky effect as a function of thermal inertia, Bond albedo, and 'uniform' surface roughness, at a fixed obliquity of 0°. For non-90° obliquity and non-zero thermal inertia the Yarkovsky effect is always enhanced by the presence of surface roughness. The Yarkovsky enhancement also increases with increasing degree of surface roughness. Fig. 8c shows the level of Yarkovsky enhancement as a function of obliquity and uniform surface roughness, and Fig. 8d shows the maximum level of Yarkovsky enhancement as a function of thermal inertia and Bond albedo. The level of enhancement gives the percentage increase in Yarkovsky semi-major axis drift for the rough surface over the smooth surface with all other asteroid properties being the same. As shown, the relative Yarkovsky enhancement increases with decreasing (but non-zero) thermal inertia and with increasing Bond albedo. At low (but non-zero) thermal inertias and high Bond albedos the relative Yarkovsky enhancement could be as much as 200% or more! However, there is only a slight increase with increasing (but non-90°) obliquity.

Yarkovsky effect predictions using 1000 realisations of each of the 'normal' and 'patchy normal' distributions produce 'rough-surface-average' predictions that exactly match the 'uniform' distribution predictions at the same degree of surface roughness. In addition, the standard deviation of the scatter of the predictions is small (i.e. less than ~3%). For example, Fig. 9a shows the percentage uncertainty in the Yarkovsky effect as a function of obliquity and surface roughness distribution type, and Fig. 9b shows the maximum percentage uncertainty as a function of thermal inertia and Bond albedo. In general, the uncertainties produced by the 'patchy normal' surface roughness distributions are about 10 times larger than those produced by the 'normal' distributions, and the 'wide' distributions produce uncertainties that are about twice as large as the 'narrow' distributions. However, all of the uncertainties are at around the few percent level or less, indicating that the degree of surface roughness is much more important than how it is distributed when predicting the Yarkovsky effect.

Unlike the Yarkovsky effect, the YORP effect is very sensitive to differences in shape for asteroids of the same size. In this case, it is not appropriate to 'shape-average' the absolute values since the YORP strength and sign can vary significantly between shape models. However, it is possible to 'shape-average' the rough surface predictions relative to their smooth surface counterparts (the sphere shape model is excluded since it produces no smooth surface YORP effect). Fig. 10a shows the absolute YORP effect dependence for the two example Gaussian-spheres shown in Fig. 6 with 'uniform' surface roughness as a function of obliquity at a fixed thermal inertia of 200 J m$^{-2}$ K$^{-1}$ s$^{-1/2}$. It demonstrates that the YORP effect is dampened by surface roughness at all obliquities where non-zero YORP torque occurs. The degree of dampening increases with increasing surface roughness. Although not shown, the degree of YORP effect dampening by surface roughness is found to be independent of thermal inertia and Bond albedo.

Fig. 10b shows the 'shape-average' degree of YORP effect dampening for the 10 different Gaussian-sphere shape models as a function of obliquity. The degree of dampening gives the percentage decrease in YORP rotational acceleration for the rough surface relative to the smooth surface with all other asteroid properties being the same. Fig. 10b shows that the degree of YORP effect dampening is least and greatest at obliquities of 90° and 0° respectively. At obliquities near the critical angle (i.e. ~55°) where zero YORP torque occurs the relative effect of surface roughness is less predictable creating the spike seen. Again it is found that the YORP effect is independent of thermal inertia in the presence of surface roughness. For these shape examples in the presence of surface roughness, the YORP effect is dampened by up to a third of the smooth surface value.

YORP effect predictions using 1000 realisations of each of the 'normal' and 'patchy normal' distributions produce 'rough-surface-average' predictions, like the Yarkovsky effect, that exactly match the 'uniform' distribution predictions at the same degree of surface roughness. However, the standard deviation of the scatter of the predictions is much larger than it is for the Yarkovsky effect. For example, Fig. 10a also shows error bars that result from the uncertainty caused by the 'wide patchy normal' surface roughness distribution on the 'medium roughness' YORP effect predictions. The size of the uncertainty can in some cases be larger than the average degree of dampening resulting in predictions that are larger in magnitude than the smooth surface case. Fig. 11 shows the relative 'shape-average' YORP effect uncertainty to the different surface roughness distribution types as a function of obliquity. It demonstrates that the relative uncertainty is roughly constant for obliquities not near the critical angle, and it is also found to be exactly constant with thermal inertia. Near the critical angle a 'random' surface roughness distribution can easily induce a YORP effect where there was none with a smooth surface, and can even produce predictions with opposite signs if there was a weak smooth surface one. Hence, the region from 40° to 60° obliquity is not plotted in Fig. 11 as the relative uncertainty in this region is very large. In general, the relative uncertainties produced by the 'patchy normal' surface roughness distributions are about 4 times larger than those produced by the 'normal' distributions, and again the 'wide' distributions produce relative uncertainties that are about twice as large as the 'narrow' distributions.

### 3.2.2 (1620) Geographos and (6489) Golevka

To test that the parameter dependencies identified in section 3.2.1 are applicable to derived shape models of real asteroids, the same investigations are applied to the light-curve-derived and radar-derived shape models of Geographos and Golevka respectively. The Geographos shape model consists of 2040 facets and the Golevka shape model consists of 4092 facets, compared with the 1152 facets used in the Gaussian-sphere shape models. The same properties listed in section 3.1 are used again for pseudo-Geographos. From the radar observations, Golevka was determined to have a size of 0.685 x 0.530 x 0.489 km with an uncertainty of 0.03 km in each dimension, a rotation period of 6.026 hours, and an orbital semi-major axis of 2.498 AU (Hudson et al. 2000). A bulk density, consistent with the Yarkovsky detection derived range, of 2700 kg m$^{-3}$ is also assumed (Chesley et al. 2003). Table 3 summarises the model parameters used for pseudo-Golevka.

Fig. 12 and Fig. 13 show the Yarkovsky effect parameter dependencies for pseudo-Geographos and pseudo-Golevka respectively, equivalent to what was shown in Fig. 8 for the Gaussian-sphere asteroids. Pseudo-Geographos shows exactly the same Yarkovsky effect parameter dependencies as the Gaussian-sphere asteroids, and pseudo-Golevka is almost as exact. The level of Yarkovsky enhancement caused by surface roughness for pseudo-Golevka (shown in Fig. 13c) is lower than that for the Gaussian-sphere asteroids and pseudo-Geographos because its larger heliocentric distance gives it a higher thermal parameter equivalent to a higher thermal inertia ($\sim$800 J m$^{-2}$ K$^{-1}$ s$^{-1/2}$ as opposed to 200 J m$^{-2}$ K$^{-1}$ s$^{-1/2}$) for the other asteroids. As shown in Fig. 8d, Fig. 12d, and Fig. 13d, at higher thermal inertias the Yarkovsky enhancement is lower which explains this result. Interestingly, Fig. 13d shows that for a Bond albedo of 0.1 and a thermal inertia >300 J m$^{-2}$ K$^{-1}$ s$^{-1/2}$ the Yarkovsky effect on pseudo-Golevka is not enhanced but is decreased very slightly by surface roughness. The thermal parameter is again equivalent to a much higher thermal inertia for the other asteroids, and the top end of the range of thermal inertias studied (i.e. 2000 J m$^{-2}$ K$^{-1}$ s$^{-1/2}$) may not have been high enough to see this effect occur for the other asteroids. Although not shown, the prediction uncertainties caused by different surface roughness distributions is again at the few percent level or less for both asteroids.

Fig. 14-15 and Fig. 16-17 show the YORP effect parameter dependencies for pseudo-Geographos and pseudo-Golevka respectively, equivalent to what was shown in Fig. 10-11 for the Gaussian-sphere asteroids. Again, pseudo-Geographos shows very similar YORP effect parameter dependencies to the Gaussian-sphere asteroids. For example, the YORP effect is dampened by surface roughness by up to a third of the smooth surface value, and the degree of dampening is greatest and least at obliquities of 0° and 90° respectively. The prediction uncertainties caused by different surface roughness distributions can also be larger than the average degree of dampening. For pseudo-Golevka, its YORP effect is very weak for an object of its size and the parameter dependencies look much different but do follow the same identified trends. The YORP effect as a function of obliquity (shown in Fig. 16a) has a much more complicated shape with zero YORP torque occurring at $\sim$10° obliquity instead of near the usual critical angle. The smooth surface predictions match those produced by Vokrouhlický & Čapek (2002) and by Breiter, Bartczak & Czekaj (2010) using the same shape model, and the complicated shape of the obliquity dependence is consistent with objects having relatively weak YORP effects (Micheli & Paolicchi 2008). As demonstrated by the error bars in Fig. 16a, the prediction uncertainties caused by different surface roughness distributions are huge in comparison to not just the average degree of dampening but also to the predictions themselves! The YORP effect in the presence of surface roughness is very unpredictable near the obliquity of zero YORP torque, and hence the relative results are not fully plotted in Fig. 16 for obliquities <20°. For about half of the obliquity range, the uncertainties caused by the 'wide patchy normal' surface roughness distributions cannot even allow the sign of the YORP effect to be accurately determined. However, taking the average of the 1000 realisations of each of the 'normal' and 'patchy normal' distributions still produce 'rough-surface-average' predictions that exactly match the 'uniform' distribution predictions at the same degree of surface roughness. As shown in Fig. 16b at 0° obliquity, the YORP effect of pseudo-Golevka can be dampened by up to one half of the smooth surface value by surface roughness.

## 4. DISCUSSION

### 4.1 Yarkovsky Effect

For predicting the Yarkovsky effect, inclusion of thermal infrared beaming caused by surface roughness can enhance the Yarkovsky orbital drift by about 200% or more in the presence of non-zero thermal inertia. The enhancement is greatest for low thermal inertias and high Bond albedos, and decreases with increasing thermal inertia and/or with decreasing Bond albedo. At very high thermal inertias and low Bond albedos the Yarkovsky orbital drift can be dampened slightly. The Yarkovsky predictions are sensitive only to the average degree of surface roughness and not to how it is distributed across an asteroid surface. This behaviour is caused by a combination of three processes linked to a rough surface undergoing selfheating previously identified in Rozitis & Green (2011):

(1) Multiple scattering of sunlight leads to a decrease in the effective Bond albedo and therefore an increase in the absorptivity of a rough surface.

(2) Re-absorption of emitted thermal radiation combined with non-zero thermal inertia results in extra-delayed thermal emission similar to an increase in thermal inertia.

(3) Thermal-infrared beaming in the presence of non-zero thermal inertia on the morning side of an asteroid is greater than that on the afternoon side.

These three processes are described in more detail here. For a rough surface with a 100% coverage of hemispherical craters the increase in sunlight absorptivity over a smooth flat surface with the same Bond albedo for individual surface elements, $\alpha$, is given by

$$\alpha = \frac{2}{2 - A_B}. \tag{33}$$

For Bond albedos of 0.1, 0.3, and 0.5 the corresponding absorptivity increases are 1.05, 1.18, and 1.33 respectively, which indicates how much extra energy can be available to fuel the Yarkovsky effect for a rough surface.

To demonstrate how extra-delayed thermal emission caused by re-absorption of emitted thermal radiation can create large increases (at low thermal inertias and high Bond albedos) and decreases (at very high thermal inertias and low Bond albedos) in the energy used to fuel the Yarkovsky effect, Fig. 18 plots the emitted thermal radiation power output as a function of rotation phase for a shape facet placed at the equator of an asteroid with a rotation period of 6 hours and an orbital semi-major axis of 1 AU. The solid lines indicate a surface with a low thermal inertia of 10 J m$^{-2}$ K$^{-1}$ s$^{-1/2}$ and a high Bond albedo of 0.5, and the dashed lines indicate a surface with a very high thermal inertia of 3000 J m$^{-2}$ K$^{-1}$ s$^{-1/2}$ and a low Bond albedo of 0.1. The thin and thick lines indicate smooth and completely rough surfaces respectively. The magnitude of the Yarkovsky effect is determined by the difference in the time-averaged power output between the asteroid afternoon (i.e. >0.5 rotation phase) and morning (i.e. <0.5 rotation phase) sides. As shown, the rough surface power output curves are rotation phase shifted compared to their smooth surface counterparts. For the smooth and rough surfaces with low thermal inertia the differences in power output between the afternoon and morning sides are 17.1 and 30.6 W m$^{-2}$ respectively, which corresponds to an increase of factor 1.79 in the energy available to fuel the Yarkovsky effect for the rough surface over its smooth surface counterpart. Furthermore, the total energy output for the rough surface divided by that for the smooth surface equals the increase in sunlight absorptivity for that Bond albedo. Similarly, for the smooth and rough surfaces with high thermal inertia the afternoon-morning differences in power output are 49.1 and 30.3 W m$^{-2}$ respectively, which corresponds to a decrease of factor 0.62 in the energy available to fuel the Yarkovsky effect. Again, the total energy output ratio between the rough and smooth surfaces is consistent with the increase in sunlight absorptivity for that Bond albedo.

Finally, since thermal-infrared beaming directs thermal emission back towards the Sun it reduces the along orbit track component of the photon recoil force. If this effect was equal across the entire asteroid surface and there were no shifts in rotational phase power output then this would lead to a decrease in Yarkovsky orbital drift. However, as demonstrated in Rozitis & Green (2011) and in Appendix B.3, the asteroid morning side beaming effect is greater than the afternoon side beaming effect. This causes the morning side along track photon force component to be reduced more than that on the afternoon side. For example, consider the surfaces described above with the Sun illuminating them at an angle of 20° above the local horizon in both the morning and afternoon. For the smooth flat surfaces, 94.0% of the overall photon force is directed parallel (afternoon side) or anti-parallel (morning side) to the orbit track regardless of the thermal inertia or Bond albedo. For the rough surface with low thermal inertia, 82.5% of the afternoon photon force is directed parallel to the orbit track, and 79.1% of the morning photon force is directed anti-parallel to the orbit track. The difference between these two values becomes larger for shape facets located at higher latitudes on the asteroid.

The net effect of these three processes described above increases for low to moderately-high thermal inertias, or decreases for very high thermal inertia, the thermal emission asymmetry between the afternoon and morning sides of an asteroid. When this asymmetry is increased it leads to an increased net photon recoil force which leads to an enhanced Yarkovsky orbital drift, and when the asymmetry is decreased it decreases the net photon recoil force which leads to a lower Yarkovsky orbital drift. As km-sized near-Earth asteroids have an average thermal inertia of $200 \pm 40$ J m$^{-2}$ K$^{-1}$ s$^{-1/2}$ (Delbo' et al. 2007), an enhancement in Yarkovsky orbital drift is mostly seen in the presence of surface roughness. If the thermal inertia is constant across an asteroid surface then the Yarkovsky enhancement/dampening contribution from each surface element in the presence of surface roughness will act in the same direction. This explains why the Yarkovsky effect is only sensitive to the average degree of surface roughness and not to how it is distributed across an asteroid surface.

## 4.2 YORP Effect

The inclusion of rough surface thermal-infrared beaming, on average, dampens the predicted YORP rotational acceleration by up to one third typically, but in some cases it can be as much as one half. The YORP effect is very sensitive to how surface roughness is distributed and predictions typically have uncertainties that are comparable to the average degree of dampening but in some cases it can be much larger. This behaviour can be explained by thermal-infrared beaming directing the emitted thermal radiation of a shape facet back towards the Sun rather than along its surface normal (as previously demonstrated in Fig. 2 and by Rozitis & Green (2011)).

For example, Fig. 19 gives a simple schematic demonstrating how thermal-infrared beaming influences the YORP torques on the example spherical asteroid with two asymmetric wedges attached that was shown in Fig. 1. The left wedge has net thermal emission at a 45° angle to the asteroid rotation axis, and the right wedge has net thermal emission at a 90° angle. This results in two opposite torques that don't cancel each other out. As the torque from the right wedge is $\sqrt{2}$ times greater in magnitude than that from the left wedge, it results in an overall torque acting in an anticlockwise sense. If both wedges were given a fully rough surface then thermal-infrared beaming directs thermal emission back towards the Sun, which increases the magnitude of both torque components. In this case, the left wedge torque is increased by a factor of 1.236, and the right wedge torque is increased by a factor of 1.079. The right wedge torque is increased less because it already directs its net smooth surface thermal emission at a maximising 90° angle to the asteroid rotation axis. For fully rough surfaces, the two opposite torques are more evenly matched resulting in lower overall torque. If the surface roughness is uniform across an asteroid surface then this would result in a dampened YORP effect.

However, if the degree of surface roughness is allowed to vary across an asteroid surface then different torque components can be increased by different amounts. As the torque components can act in opposite senses, the varying degree of changes in torques can cause the YORP effect to be dampened even further or enhanced. For example, if only the left wedge in Fig. 19 was rough then the YORP effect is dampened more than if both wedges were rough. If only the right wedge was rough then the YORP effect is enhanced, but only slightly, compared to if both wedges were smooth. This explains why the YORP effect is so sensitive to how the surface roughness is distributed, especially so in extreme cases where one side of the asteroid is smooth and the other side is completely rough.

The YORP rotational acceleration is still independent of thermal inertia in the presence of surface roughness. Since thermal inertia still does not affect the total amount of energy thermally reprocessed at a given latitude of an asteroid in the presence of surface roughness, the explanation of thermal inertia independence given by Čapek & Vokrouhlický (2004) is still valid. It also remains independent of Bond albedo. However, the degree of surface roughness produces different fractions of YORP torque arising from the reflected sunlight and thermally emitted components for rough surfaces compared to those of smooth surfaces with the same Bond albedo. For example, Table 4 lists the average fractions of YORP torque components at different Bond albedos and degrees of surface roughness for the shape examples studied here at 0° obliquity.

### 4.3 Implications of Rough Surface Thermal-Infrared Beaming

The influence of rough surface thermal-infrared beaming effects on the YY predictions investigated in section 3.2 on shape models representative of asteroids reveal that they are as important as the conventional parameters discussed in section 3.1.

Since rough surface thermal-infrared beaming can enhance the Yarkovsky effect, it means that any asteroid mass that has been inferred by comparing the observed orbital drift with a smooth surface model prediction would be underestimated, as would the associated bulk density. This could be the case for Golevka whose mass and bulk density was determined by this method (Chesley et al. 2003). The enhancement could also pose problems for accurately predicting the close encounters with Earth of potentially hazardous asteroids under the influence of the Yarkovsky effect e.g. Apophis (Giorgini et al. 2008). In both cases, more reliable estimates should involve producing a Yarkovsky prediction using thermal inertia and surface roughness values that are consistent with those inferred from thermal-infrared observations.

If the example asteroids studied here are normalised so that they have the same size, bulk density, and semi-major axis, then the YORP rotational acceleration prediction uncertainties, caused by different surface roughness distributions, can be studied as a function of the magnitude of the smooth surface YORP rotational acceleration. Fig. 20 shows this relationship for the example asteroids under the influence of the 'wide patchy normal' surface roughness distributions at 0° and 90° obliquity. The asteroids have been normalised to have a volume equivalent to a sphere 1 km in diameter, a bulk density of 2500 kg m$^{-3}$, and a semi-major axis of 1 AU. As shown, at both obliquities the YORP effect prediction uncertainties follow an approximate inverse relationship with the magnitude of the smooth surface YORP rotational acceleration. This implies that weaker YORP effect predictions are more susceptible to how surface roughness is distributed. This sounds plausible when considering an extreme case where surface roughness variations are added to a sphere to induce a YORP rotational acceleration, which is zero when the sphere is completely smooth, and causes an infinite relative change. This relationship requires further investigation by increasing the number of example asteroids studied but this is beyond the scope of this paper. However, this approximate relationship could be used to estimate the effect of surface roughness on the YORP effect predictions for other asteroids. For example, an obvious choice is to estimate what influence it has on the YORP effect predictions for asteroid (25143) Itokawa. YORP effect modelling of Itokawa using the Hayabusa derived shape models in the absence of surface roughness predict a rotational deceleration that should be easily observable (e.g. Scheeres et al. 2007; Breiter et al. 2009). However, light curve observations of Itokawa fail to see any change in rotation rate, which contradicts the predictions (Ďurech et al. 2008a). Since Itokawa has an almost 180° obliquity (i.e. equivalent to 0° obliquity but with retrograde rotation) this allows the relationship shown in Fig. 20 to be directly applicable once Itokawa's YORP effect predictions have been normalised. Using the work of Breiter et al. (2009), Itokawa's normalised YORP rotational acceleration predictions range from -1.9 to -11.2 x10$^{-3}$ rad yr$^{-2}$ for the different variants of its shape model, which leads to relative prediction uncertainties ranging from 310% to 50% respectively. This suggests that Itokawa's shape needs to be known to at least 1-cm scales (i.e. the size of its thermal skin depth) in order to make an accurate YORP effect prediction.

At present, there is insufficient information to see whether including the effects of rough surface thermal-infrared beaming would reduce or increase the apparently high sensitivity of the YORP effect to small-scale variations in an asteroid shape model, e.g. as demonstrated by Breiter et al. (2009) and Statler (2009). To properly investigate this, thermal-infrared beaming would need to be combined with global-selfheating effects, which could also play a key role. The shape studies of Breiter et al. (2009) and Statler (2009) should then be repeated whilst including the combined effects of thermal-infrared beaming and global-selfheating. This is a subject for future work, and the implementation and influence of global-selfheating on the Yarkovsky and YORP effects will be presented in detail in a future paper.

## 5. SUMMARY AND CONCLUSIONS

The *ATPM* presented in Rozitis & Green (2011) has been adapted to simultaneously predict the Yarkovsky and YORP effects acting on an asteroid and includes rough surface thermal-infrared beaming effects. This is the first such model of its kind, and a detailed investigation into the influence of these effects on several shapes representative of asteroids reveals that surface roughness is as important as the more conventional parameters used in previous models.

The Yarkovsky effect is found to be not highly sensitive to shape for asteroids of the same size. For non-zero thermal inertia and non-90° obliquity, the Yarkovsky semi-major axis drift is always enhanced by the presence of surface roughness, except for extremely large values of the thermal parameter (e.g. $\Theta > 10$). The degree of enhancement increases with increasing surface roughness, and is typically tens of percent but can be as much as a factor of 2. The relative enhancement also increases with decreasing thermal inertia and increasing Bond albedo, but is not sensitive to obliquity. The degree of semi-major axis drift is not sensitive to the distribution of surface roughness, and the average degree of surface roughness is therefore much more important than how it is distributed.

In contrast, the YORP effect is very sensitive to differences in shape. However, the magnitude of the YORP rotational acceleration is on average dampened by surface roughness for all obliquities, and the degree of dampening increases with increasing surface roughness. The degree of dampening is independent of the thermal inertia or Bond albedo. For Gaussian-sphere shape models the degree of dampening can be up to a third of the smooth surface prediction, but it can be more for more extreme shapes (e.g. up to one half for the radar-derived shape model of Golevka). Unlike the Yarkovsky effect, the YORP rotational acceleration is sensitive to the spatial distribution of roughness over the asteroid surface.

*ATPM* is the first model which incorporates thermal-infrared beaming for interpretation of thermal-infrared observations and determination of the associated Yarkovsky and YORP effects for the derived surface thermal properties. The model can be used to improve predictions of Yarkovsky orbital drift, which has consequences for asteroid mass and bulk density determination from observed orbital drift, and for close encounter predictions of potentially hazardous asteroids. However, for predicting the YORP rotational acceleration, the sensitivity to detailed shape and surface roughness mean that the predictions are only likely to be reliable when a very detailed shape model (such as those obtained from spacecraft imaging or from very high resolution radar observations) is available, and when some indication of the distribution of surface roughness is also available. If these data are not available, then the rough surface model can be used to obtain a realistic indication of the range of possible YORP rotational accelerations.

Finally, since *ATPM* also includes global-selfheating effects, their influence on the Yarkovsky and YORP effects will be studied in detail in a future paper.

**Acknowledgments**

We are grateful to the reviewer S. Breiter for several suggested refinements to the manuscript. We are also grateful to the anonymous referee who reviewed a previous version of this manuscript, and gave several comments and suggestions which led to this significantly improved version. The work of BR is supported by the UK Science and Technology Facilities Council (STFC).

**References**

Bottke Jr. W. F., Vokrouhlický D., Rubincam D. P., Nesvorný D., 2006, Ann. Rev. Earth Planet. Sci., 34, 157

Breiter S., Bartczak P., Czekaj M., 2010, MNRAS, 408, 1576

Breiter S., Bartczak P., Czekaj M., Oczujda B., Vokrouhlický D., 2009, A&A, 507, 1073

Breiter S., Michalska H., 2008, MNRAS, 388, 927

Breiter S., Michalska H., Vokrouhlický D., Borczyk W., 2007, A&A, 471, 345

Breiter S., Rożek A., Vokrouhlický D., 2011, MNRAS, 417, 2478

Breiter S., Vokrouhlický D., 2011, MNRAS, 410, 2807

Breiter S., Vokrouhlický D., Nesvorný D., 2010, MNRAS, 401, 1933

Brož M., 2006, PhD thesis, Charles University

Čapek D., 2007, PhD thesis, Charles University

Čapek D., Vokrouhlický D., 2004, Icarus, 172, 526

Chesley S. R., et al., 2003, Science, 302, 1739

Chesley S. R., Vokrouhlický D., Ostro S. J., Benner L. A. M., Margot J.-L., Matson R. L., Nolan M. C., Shepard M. K., 2008, in Asteroids, Comets, Meteors, LPI Contr. 1405, 8330

Cicalò S., Scheeres D. J., 2010, Celestial Mechanics and Dynamical Astronomy, 106, 301

Delbo' M., dell'Oro A., Harris A. W., Mottola S., Mueller M., 2007, Icarus, 190, 236

Delbo' M., Tanga P., 2009, P&SS, 57, 259

Ďurech J., et al., 2008a, A&A, 488, 345

Ďurech J., et al., 2008b, A&A, 489, L25

Ďurech J., et al., 2009, in AAS/Division for Planetary Sciences Meeting Abstracts Vol. 41, #56.04

Giorgini J. D., Benner L. A. M., Ostro S. J., Nolan M. C., Busch M. W., 2008, Icarus, 193, 1

Groussin O., et al., 2007, Icarus, 187, 16

Hapke B., 2002, Icarus, 157, 523

Holsapple K. A., 2010, Icarus, 205, 430

Hudson R. S., Ostro S. J., 1999, Icarus, 140, 369

Hudson R. S., et al., 2000, Icarus, 148, 37

Jakosky B. M., Finiol G. W., Henderson B. G., 1990, Geophys. Res. Lett., 17, 985

Kaasalainen M., Ďurech J., Warner B. D., Krugly Y. N., Gaftonyuk N. M., 2007, Nature, 446, 420

Lagerros J. S. V., 1998, A&A, 332, 1123

Lowry S. C., et al., 2007, Science, 316, 272

Micheli M., Paolicchi P., 2008, A&A, 490, 387

Muinonen K., Lagerros J. S. V., 1998, A&A, 333, 753

Müller T. G., Lagerros J. S. V., 1998, A&A, 338, 340

Mysen E., 2008a, A&A, 484, 563

Mysen E., 2008b, MNRAS, 383, L50

Nesvorný D., Bottke W. F., 2004, Icarus, 170, 324

Nesvorný D., Vokrouhlický D., 2007, AJ, 134, 1750
Nesvorný D., Vokrouhlický D., 2008, A&A, 480, 1
Ostro S. J., et al., 2006, Science, 314, 1276
Pravec P., Harris A. W., 2007, Icarus, 190, 250
Pravec P., et al., 2010, Nature, 466, 1085
Rozitis B., Green S. F., 2011, MNRAS, 415, 2042
Rubincam D. P., 2000, Icarus, 148, 2
Rubincam D. P., 2007, Icarus, 192, 460
Rubincam D. P., Paddack S. J., 2010, Icarus, 209, 863
Saari J. M., Shorthill R. W., Winter D. F., 1972, The Moon, 5, 179
Scheeres D. J., 2007, Icarus, 188, 430
Scheeres D. J., Abe M., Yoshikawa M., Nakamura R., Gaskell R. W., Abell P. A., 2007,
    Icarus, 188, 425
Scheeres D. J., Mirrahimi S., 2008, Celestial Mechanics and Dynamical Astronomy, 101, 69
Sekiya M., Shimoda A. A., Wakita S., 2012, PSS, 60, 304
Spencer J. R., 1990, Icarus, 83, 27
Spencer J. R., Lebofsky L. A., Sykes M. V., 1989, Icarus, 78, 337
Spitale J., Greenberg R., 2001, Icarus, 149, 222
Statler T. S., 2009, Icarus, 202, 502
Steinberg E., Sari R., 2011, AJ, 141, 55
Taylor P. A., et al., 2007, Science, 316, 274
Vokrouhlický D., 2006, A&A, 459, 275
Vokrouhlický D., Čapek D., 2002, Icarus, 159, 449
Vokrouhlický D., Čapek D., Chesley S. R., Ostro S. J., 2005, Icarus, 179, 128
Vokrouhlický D., Chesley S. R., Matson R. D., 2008, AJ, 135, 2336
Vokrouhlický D., Farinella P., 1998, AJ, 116, 2032
Vokrouhlický D., Milani A., Chesley S. R., 2000, Icarus, 148, 118
Vokrouhlický D., Nesvorný D., Bottke W. F., 2003, Nature, 425, 147
Walsh K. J., Richardson D. C., Michel P., 2008, Nature, 454, 188
Žižka J., Vokrouhlický D., 2011, Icarus, 211, 511

**Appendix A: Gaussian Random Sphere Shape Models**

The Gaussian random sphere shape models used here are produced using the formalism outlined in Muinonen & Lagerros (1998). The radial distance from the origin, $r(\theta,\varphi)$, is given by

$$r\left(\theta,\varphi\right) = \exp\left[s\left(\theta,\varphi\right) - \frac{\beta^2}{2}\right] \tag{A1}$$

where $(\theta,\varphi)$ are the spherical-coordinate angles, and $s(\theta,\varphi)$ is a function called the log-radius which has standard deviation $\beta$. The log-radius is defined by an expansion in spherical harmonics given by

$$s\left(\theta,\varphi\right) = \sum_{l=0}^{\infty}\sum_{m=-l}^{l} s_{lm} Y_{lm}\left(\theta,\varphi\right) \tag{A2}$$

where $Y_{lm}(\theta,\varphi)$ are the orthonormal spherical harmonics with Condon-Shortley phase. Since the log-radius is real, the spherical harmonics coefficients $s_{lm}$ are constrained by the relation

$$s_{l,-m} = \left(-1\right)^m s_{lm}^* \ . \tag{A3}$$

The real and imaginary parts of the spherical harmonics coefficients are independent Gaussian random variables with zero means and variances given by

$$\text{Var}\big[\text{Re}\big(s_{lm}\big)\big] = \big(1 + \delta_{m0}\big)\frac{2\pi}{2l+1}C_l \tag{A4}$$

$$\text{Var}\big[\text{Im}\big(s_{lm}\big)\big] = \big(1 - \delta_{m0}\big)\frac{2\pi}{2l+1}C_l \tag{A5}$$

where $\delta_{ij}$ is the Kronecker delta symbol, and $C_l$ are the Legendre coefficients of the log-radius covariance function. The Legendre coefficients also define the log-radius standard deviation by

$$\beta^2 = \sum_{l=0}^{\infty} C_l . \tag{A6}$$

Muinonen & Lagerros (1998) estimate the Legendre coefficients for $l \leq 10$ from shapes derived from light curve observations of 14 objects, and Statler (2009) adopts an analytic fit to their results to give

$$C_l = 1.2\frac{\big(l^2 + 0.26\big)^2}{\big(l^8 + 90.0\big)^{1.06}} . \tag{A7}$$

This analytic fit is used to extrapolate the covariance function out to $l = 20$, which Statler (2009) states is required to achieve a shape with a 10% YORP prediction uncertainty.

 To generate the Gaussian random sphere shape models, vertices are spaced out equally by 7.5° in the spherical-coordinate system space, which leads to shape models consisting of 578 vertices and 1152 facets. The radial distance of each of these vertices is produced with the method outlined above using independent realisations of the Gaussian random spherical harmonic coefficients for each shape model. Assuming a uniform internal density distribution, the models are shifted so that the centre of mass lies at the coordinate system origin, and rotated so that the vector of maximum moment of inertia lies parallel with the coordinate system z axis.

## Appendix B: Rozitis & Green (2011) Errata

### B.1 Total Incident Thermal Flux

The $(1 - A_{\text{TH}})$ term given in equation 21 of Rozitis & Green (2011), which is used for the calculation of total incident thermal flux from interfacing facets, is not required since it is already included in equations 1 and 6. It is used correctly in the equivalent equations given in this paper.

### B.2 Planck Function Emissivity

As the emissivity $\varepsilon$ is included in equation 1 of Rozitis & Green (2011), which is used for the calculation of surface temperatures, it should also be included as a multiplying factor in the Planck function given by equation 22.

### B.3 Figure 6e

Unfortunately, Fig. 6e of Rozitis & Green (2011) does not show the directionally resolved dependence of total radiated power integrated over all wavelengths as a function of different Sun illumination angles. Instead, it shows a similar looking figure that demonstrates the directionally resolved dependence of the overall thermal-infrared beaming effect caused by macroscopic roughness combined with microscopic roughness. To simulate microscopic beaming, the thermal emission from each facet is described by a $\cos^n \theta$ law rather than by the usual Lambert emission law. The figure shows the directionally resolved dependence of 10 μm thermal emission at Sun illumination angles of ±70° for a rough surface placed on the equator of an example asteroid with different values of $n$. The example asteroid is placed at 1 AU from the Sun and has a rotation period 6 hours, a Bond albedo of 0.1, and a thermal inertia of 200 J m$^{-2}$ K$^{-1}$ s$^{-1/2}$. The rough surface consists of a 50% coverage of hemispherical craters. The solid, dashed, and dotted lines in the figure correspond to $n$ values of 1, 2, and 3 respectively. It shows that the maximum overall thermal-infrared beaming effect is only dependent on macroscopic roughness and not microscopic roughness, which was in agreement with findings of other works identified in that paper and was not meant to be included. The correct figure that was meant to be included is shown in Fig. B1.

**Tables**

*Table 1: Summary of Yarkovsky and YORP models produced over the past decade.*

| Model* | Yarkovsky / YORP | Thermal Model Type | Shape Type | Projected Shadows | Other Features |
|--------|------------------|--------------------|------------|-------------------|----------------|
| *ATPM* | Yarkovsky & YORP | 1D or Instantaneous | Polyhedral | Yes | Corrected Emission Vectors, Thermal-Infrared Beaming & Global-Selfheating** |
| Sekiya, Shimoda & Wakita 2012 | Yarkovsky | 3D | Sphere | N/A | |
| Breiter, Rożek, & Vokrouhlický 2011 | YORP | Instantaneous | Polyhedral | Yes | |
| Steinberg & Sari 2011 | Binary YORP | Instantaneous or Rotation Averaged | Polyhedral | No | |
| Breiter & Vokrouhlický 2011 | YORP | 1D | Polyhedral | Yes | Microscopic Beaming |
| Breiter, Bartczak & Czekaj 2010 | YORP | 1D | Polyhedral | Yes | Depth Dependent Properties |
| Cicalò & Scheeres 2010 | YORP | Instantaneous | Polyhedral | Yes | |
| Breiter, Vokrouhlický & Nesvorný 2010 | YORP | Linearised 3D | Spherical Harmonics | No | |
| Breiter et al. 2009 | YORP | Instantaneous | Polyhedral | Yes | |
| Statler 2009 | YORP | Instantaneous | Polyhedral | Yes | Corrected Emission Vectors |
| Micheli & Paolicchi 2008 | YORP | Instantaneous | Polyhedral | Yes | |
| Chesley et al. 2008 | Yarkovsky | Inverse Square-Law | N/A | N/A | |
| Mysen 2008b | Yarkovsky & YORP | Linearised 1D | Polyhedral | Yes | |
| Breiter & Michalska 2008 | YORP | Linearised 1D | Spherical Harmonics | No | |
| Scheeres & Mirrahimi 2008 | YORP | Time Delay | Polyhedral | Yes | |
| Mysen 2008a | Yarkovsky & YORP | Time Delay | Polyhedral | Yes | |
| Čapek 2007 | Yarkovsky & YORP | 1D | Polyhedral | Yes | Temperature & Depth Dependent Properties |
| Nesvorný & Vokrouhlický 2007 | YORP | Linearised 1D | Spherical Harmonics | No | |
| Breiter et al. 2007 | YORP | Instantaneous or Time Delay | Spheroid | N/A | |
| Kaasalainen et al. 2007 | YORP | Instantaneous | Polyhedral | N/A | |
| Scheeres 2007 | YORP | Time Delay | Polyhedral | No | |
| Brož 2006 | Yarkovsky | 1D or Linearised 1D | Sphere | N/A | |
| Vokrouhlický 2006 | Yarkovsky | Linearised 1D | Sphere | N/A | Variable Albedo |
| Vokrouhlický et al. 2005 | Yarkovsky on Binaries | 1D | Polyhedral | Yes | |
| Čapek & Vokrouhlický 2004 | YORP | 1D | Polyhedral | Yes | |
| Chesley et al. 2003 | Yarkovsky | 1D | Polyhedral | Yes | |
| Vokrouhlický & Čapek 2002 | YORP | Instantaneous | Polyhedral | Yes | |
| Spitale & Greenberg 2001 | Yarkovsky | 3D | Sphere | N/A | |
| Vokrouhlický, Milani & Chesley 2000 | Yarkovsky | Linearised 1D | Sphere | N/A | |
| Rubincam 2000 | YORP | Instantaneous | Spherical Harmonics | Yes | |

*Only publications describing the first implementation of the models is listed to avoid duplicate entries.
**The ATPM global-selfheating will be presented in detail in a future paper.*

*Table 2: Surface roughness distribution properties.*

| Distribution | Degree of Surface Roughness | Roughness Fraction $f_R$ | RMS Slope / ° |
|---|---|---|---|
| 'Uniform' | 'Zero' | 0.00 | 0.0 |
| | 'Low' | 0.25 | 24.8 |
| | 'Medium' | 0.50 | 35.1 |
| | 'High' | 0.75 | 43.0 |
| | 'Full' | 1.00 | 49.6 |
| 'Normal' & 'Patchy Normal' | 'Low Narrow' | 0.25 ± 0.08 | $24.8 \, ^{+3.7}/_{-4.3}$ |
| | 'Medium Narrow' | 0.50 ± 0.08 | $35.1 \, ^{+2.7}/_{-3.0}$ |
| | 'Medium Wide' | 0.50 ± 0.16 | $35.1 \, ^{+5.2}/_{-6.2}$ |
| | 'High Narrow' | 0.75 ± 0.08 | $43.0 \, ^{+2.2}/_{-2.4}$ |

*Table 3: Asteroid model parameters.*

| Parameter | Pseudo-Geographos | Gaussian Random Spheres | Pseudo-Golevka |
|---|---|---|---|
| Number of vertices | 1022 | 578 | 2048 |
| Number of facets | 2040 | 1152 | 4092 |
| Diameter of equivalent volume sphere | 2.56 km | 1.00 km | 0.52 km |
| Bulk density | 2500 kg m$^{-3}$ | 2500 kg m$^{-3}$ | 2700 kg m$^{-3}$ |
| Mass | 2.2x10$^{13}$ kg | 1.3x10$^{12}$ kg | 2.0x10$^{11}$ kg |
| Moment of inertia | 3.4x10$^{19}$ kg m$^2$ | (1.7-2.4)x10$^{17}$ kg m$^2$ | 7.1x10$^{15}$ kg m$^2$ |
| Semi-major axis | 1.246 AU | 1.000 AU | 2.498 AU |
| Rotation period | 5.223 hours | 6.000 hours | 6.026 hours |

*Table 4: YORP torque components.*

| | | Degree of Surface Roughness | | | | |
|---|---|---|---|---|---|---|
| | | 'Zero' | 'Low' | 'Medium' | 'High' | 'Full' |
| **Bond Albedo** | 0.1 | 90.0% Th. 10.0% Ref. | 91.3% Th. 8.7% Ref. | 92.9% Th. 7.1% Ref. | 94.9% Th. 5.1% Ref. | 97.4% Th. 2.6% Ref. |
| | 0.3 | 70.0% Th. 30.0% Ref. | 73.4% Th. 26.6% Ref. | 77.4% Th. 22.6% Ref. | 82.4% Th. 17.6% Ref. | 88.7% Th. 11.3% Ref. |
| | 0.5 | 50.0% Th. 50.0% Ref. | 54.5% Th. 45.5% Ref. | 60.0% Th. 40.0% Ref. | 66.7% Th. 33.3% Ref. | 75.1% Th. 24.9% Ref. |

*Th. - Thermal component, Ref. - Reflected component

**Figure Captions**

Figure 1: Schematic of the Yarkovsky and YORP effects on the orbit and spin properties of a small asteroid.

Figure 2: Schematic of the rough surface thermal-infrared beaming effect.

Figure 3: Schematic of the Advanced Thermophysical Model (*ATPM*) where the terms $F_{\text{SUN}}$, $F_{\text{SCAT}}$, $F_{\text{RAD}}$, $k_c(\text{d}T/\text{d}x)$, and $\varepsilon\sigma T^4$ are the direct sunlight, multiple scattered sunlight, reabsorbed thermal radiation, conducted heat, and thermal radiation lost to space respectively (copied from Rozitis & Green (2011)).

Figure 4: Shape model of the hemispherical crater used to induce rough surface thermal-infrared beaming.

Figure 5: Profiles used to create the different types of surface roughness distribution. The different line styles correspond to the degrees of surface roughness indicated by the legend at the top. The thin lines represent the 'narrow normal' distributions, and the thick line represents the 'wide normal' distribution.

Figure 6: Shapes and typical surface temperature distributions of two Gaussian random sphere asteroids (top), and of asteroids Geographos (bottom left) and Golevka (bottom right). Their sizes are not drawn to scale. The vertical lines correspond to the rotation poles and the diagonal lines point towards the Sun. The hottest facets are coloured white and the coldest are coloured black with a transition between the two for facets of intermediate temperatures.

Figure 7: Parameter sensitivity of the Yarkovsky and YORP effects acting on a smooth pseudo-Geographos. (a) Yarkovsky orbital drift as a function of thermal inertia (x axis) and Bond albedo (legend). (b) Yarkovsky orbital drift as a function of thermal inertia (x axis) and heliocentric distance (legend given in AU). (c) Diurnal and seasonal Yarkovsky orbital drift as a function of obliquity. The seasonal effect has been multiplied by a factor of -100 so that it can be plotted on the same axes as the diurnal effect. (d) YORP rotational acceleration as a function of obliquity.

Figure 8: Surface roughness parameter sensitivity of the 'shape-average' Yarkovsky effect acting on Gaussian random sphere asteroids. (a) Yarkovsky orbital drift as a function of obliquity (x axis) and surface roughness (legend). (b) Yarkovsky orbital drift as a function of thermal inertia (x axis) and Bond albedo (legend) in the presence of 'zero' (thin lines) and 'full' (thick lines) surface roughness. (c) Enhancement of the Yarkovsky orbital drift as a function of obliquity (x axis) and surface roughness (legend of panel (a)). (d) Maximum enhancement of the Yarkovsky orbital drift for 'full' surface roughness as a function of thermal inertia (x axis) and Bond albedo (legend of panel (b)).

Figure 9: Uncertainty of the 'shape-average' Yarkovsky effect acting on Gaussian random sphere asteroids caused by different surface roughness distributions. (a) Uncertainty of the Yarkovsky orbital drift as a function of obliquity (x axis) and surface roughness distribution type. The solid, long-dashed, dashed-dotted, and short-dashed lines coresspond to the 'narrow normal', 'wide normal', 'narrow patchy normal', and 'wide patchy normal' surface roughness distributions respectively. (b) Maximum uncertainty of the Yarkovsky orbital drift caused by the 'wide patchy normal' surface roughness distribution as a function of thermal inertia (x axis) and Bond albedo (legend).

Figure 10: Surface roughness parameter sensitivity of the YORP effect acting on Gaussian random sphere asteroids. (a) YORP rotational acceleration as a function of obliquity (x axis) and surface roughness (legend) for two example asteroids. (b) 'Shape-average' dampening of the YORP rotational acceleration as a function of obliquity (x axis) and surface roughness (legend and legend of panel (a)).

Figure 11: Uncertainty of the 'shape-average' YORP rotational acceleration acting on Gaussian random sphere asteroids caused by different surface roughness distributions as a function of obliquity (x axis). The solid, long-dashed, dashed-dotted, and short-dashed lines of both panels corespond to the 'narrow normal', 'wide normal', 'narrow patchy normal', and 'wide patchy normal' surface roughness distributions respectively.

Figure 12: Surface roughness parameter sensitivity of the Yarkovsky effect acting on pseudo-Geographos. (a) Yarkovsky orbital drift as a function of obliquity (x axis) and surface roughness (legend). (b) Yarkovsky orbital drift as a function of thermal inertia (x axis) and Bond albedo (legend) in the presence of 'zero' (thin lines) and 'full' (thick lines) surface roughness. (c) Enhancement of the Yarkovsky orbital drift as a function of obliquity (x axis) and surface roughness (legend of panel (a)). (d) Maximum enhancement of the Yarkovsky orbital drift for full surface roughness as a function of thermal inertia (x axis) and Bond albedo (legend of panel (b)).

Figure 13: Surface roughness parameter sensitivity of the Yarkovsky effect acting on pseudo-Golevka. Same as Fig. 12 except for pseudo-Golevka.

Figure 14: Surface roughness parameter sensitivity of the YORP effect acting on pseudo-Geographos. (a) YORP rotational acceleration as a function of obliquity (x axis) and surface roughness (legend). (b) 'Shape-average' dampening of the YORP rotational acceleration as a function of obliquity (x axis) and surface roughness (legend and legend of panel (a)).

Figure 15: Uncertainty of the YORP rotational acceleration acting on pseudo-Geographos caused by different surface roughness distributions as a function of obliquity (x axis). The solid, long-dashed, dashed-dotted, and short-dashed lines of both panels corespond to the 'narrow normal', 'wide normal', 'narrow patchy normal', and 'wide patchy normal' surface roughness distributions respectively.

Figure 16: Surface roughness parameter sensitivity of the YORP effect acting on pseudo-Golveka. Same as Fig. 14 except for pseudo-Golevka, and where the circle, square, diamond, and triangle markers indicate the 0° obliquity values for low, medium, high, and full surface roughness respectively.

Figure 17: Uncertainty of the YORP effect acting on pseudo-Golevka caused by different surface roughness distributions. Same as Fig. 15 except for pseudo-Golevka, and where the triangle, diamond, square, and circle markers indicate the 0° obliquity values for the 'narrow normal', 'wide normal', 'narrow patchy normal', and 'wide patchy normal' surface roughness distributions respectively.

Figure 18: Power output as a function of rotation phase for a surface element located on the equator of an example asteroid. The solid lines represent a surface with Bond albedo and thermal inertia of 0.5 and 10 J m$^{-2}$ K$^{-1}$ s$^{-1/2}$ respectively, and the dashed lines represent a surface with equivalent values of 0.1 and 3000 J m$^{-2}$ K$^{-1}$ s$^{-1/2}$. The thin and thick lines correspond to smooth and completely rough surfaces respectively.

Figure 19: The influence of rough surface thermal-infrared beaming on YORP torques. The wedges shown correspond to those attached to the example spherical asteroid shown in Fig. 1.

Figure 20: YORP rotational acceleration prediction uncertainty as a function of normalised (to 1 km diameter, 2500 kg m$^{-3}$ bulk density, and 1 AU orbital semi-major axis) rotational acceleration at 0° (solid markers) and 90° (open markers) obliquity. The circles, triangle, and square correspond to the Gaussian sphere, Geographos, and Golevka shape models respectively. The lines are the best power law fits to the trends indicated, and the equations and R-squared values of the fits are also given next to them.

Figure B1: Directionally resolved dependence of total radiated power integrated over all wavelengths as a function of different Sun illumination angles (legend) for a rough asteroid surface. This was meant to be Fig. 6e of Rozitis & Green (2011).

**Figures**

Figure 1:

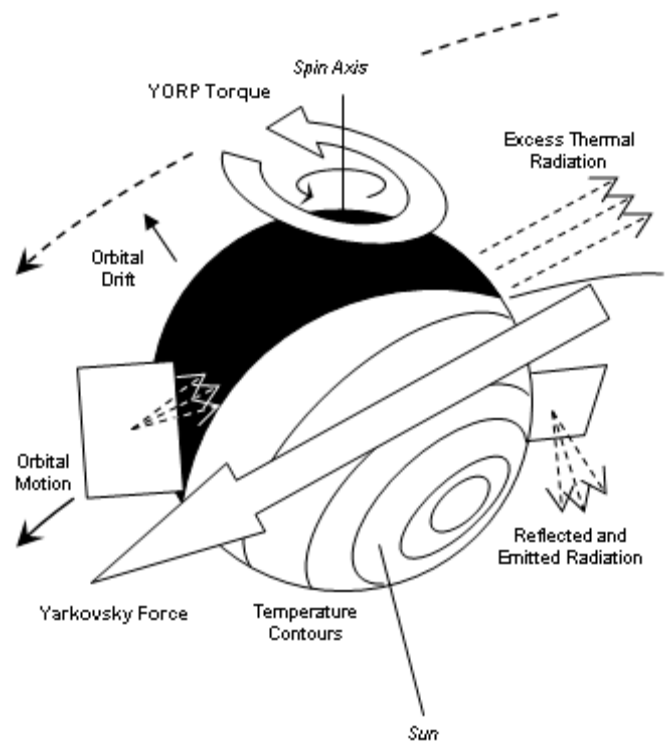

Figure 2:

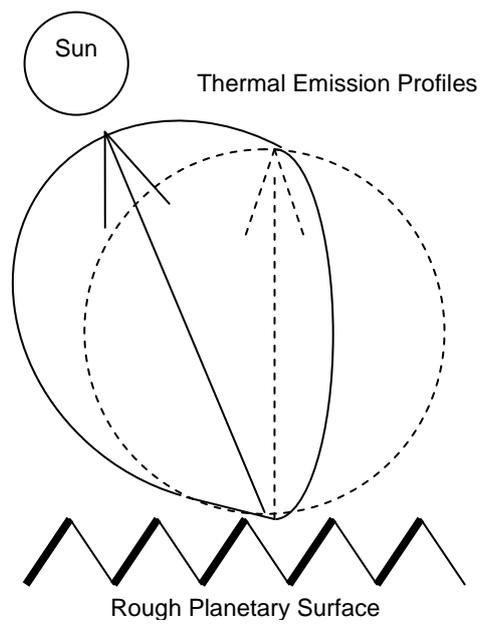

Figure 3:

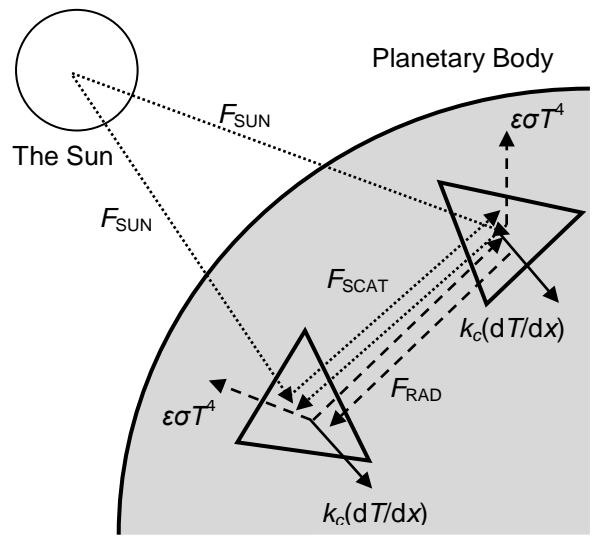

Figure 4:

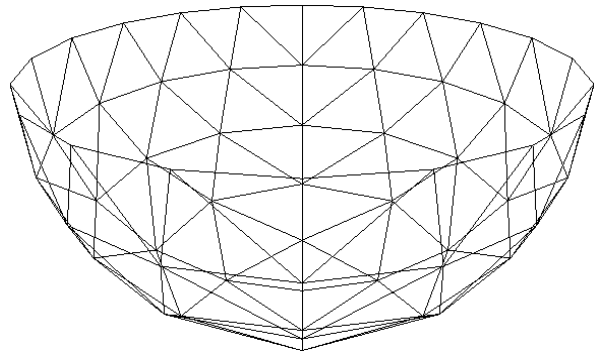

Figure 5:

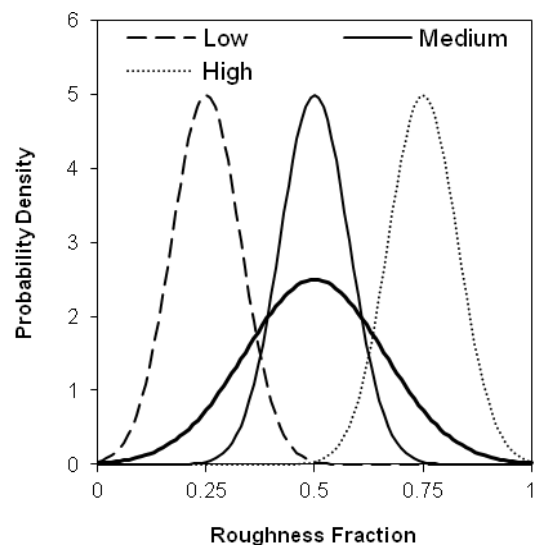

Figure 6:

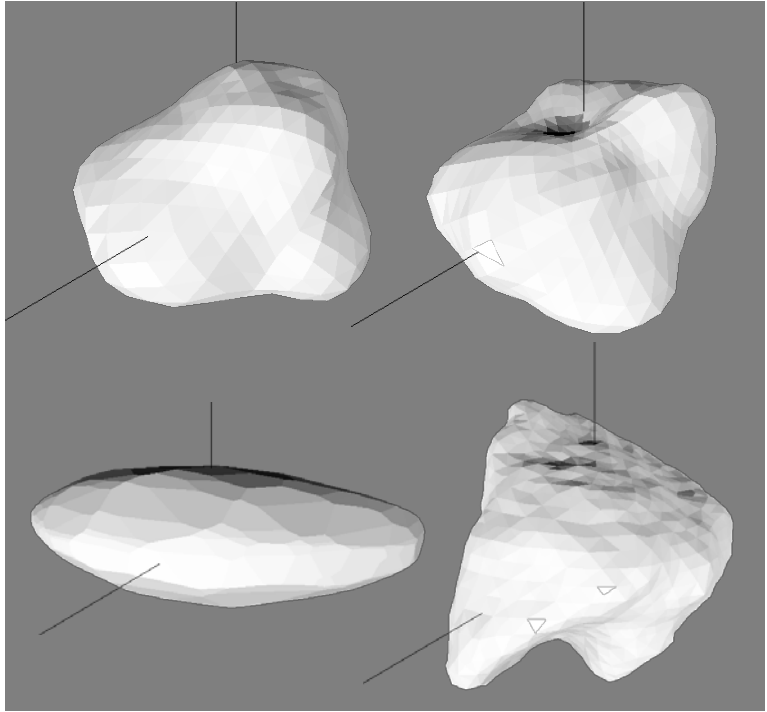

Figure 7:

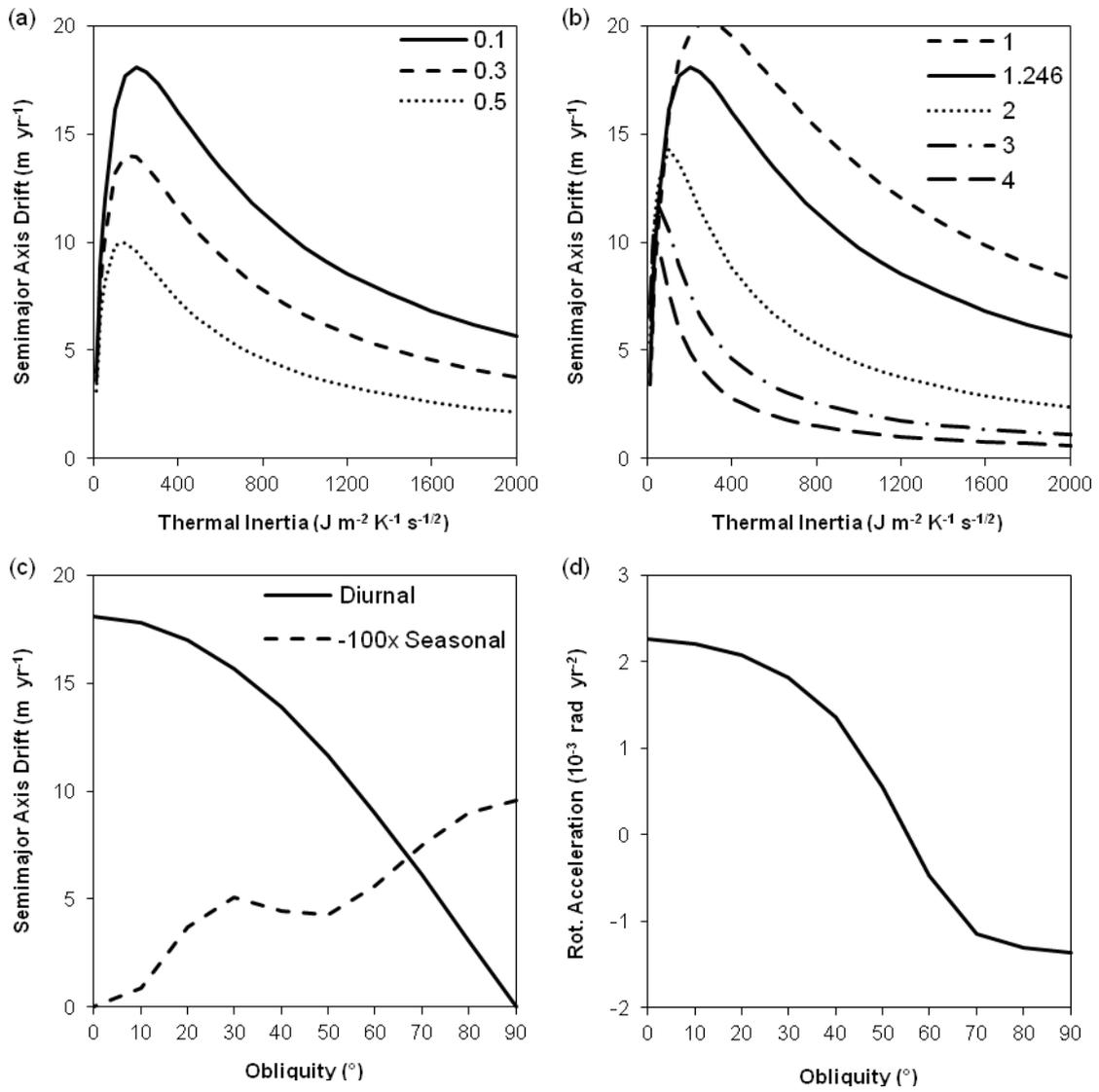

Figure 8:

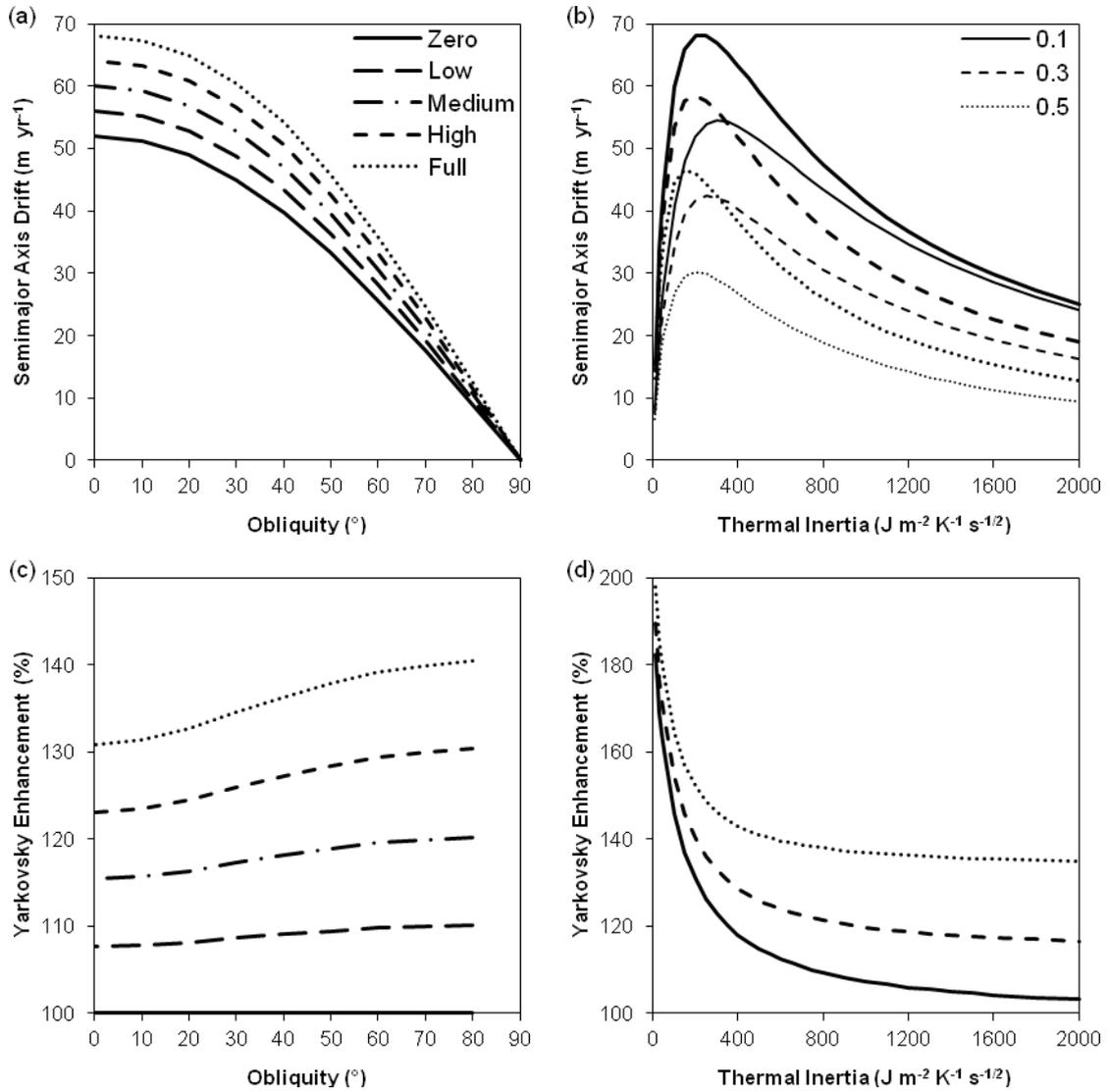

Figure 9:

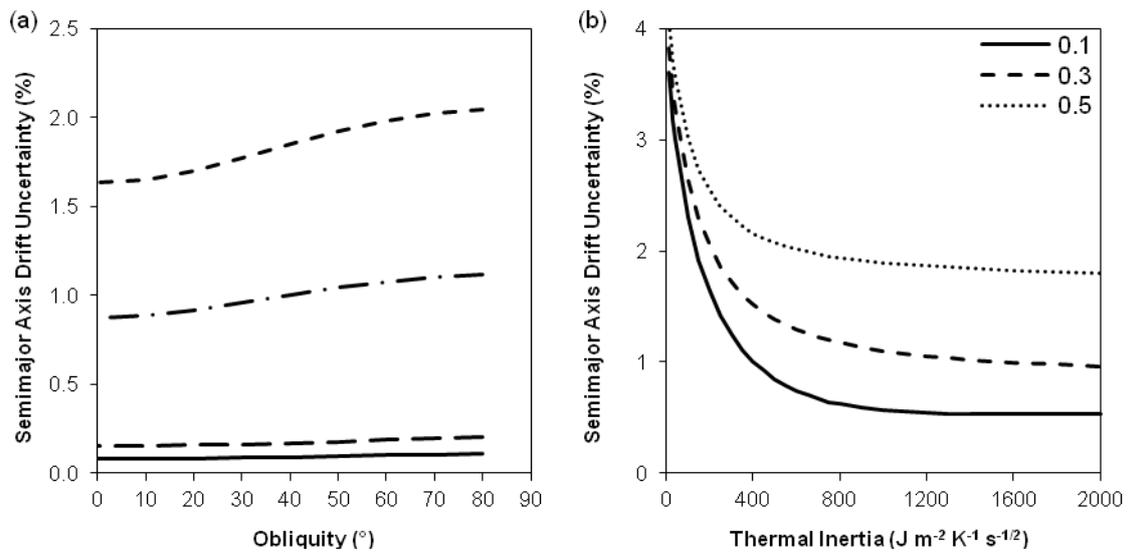

Figure 10:

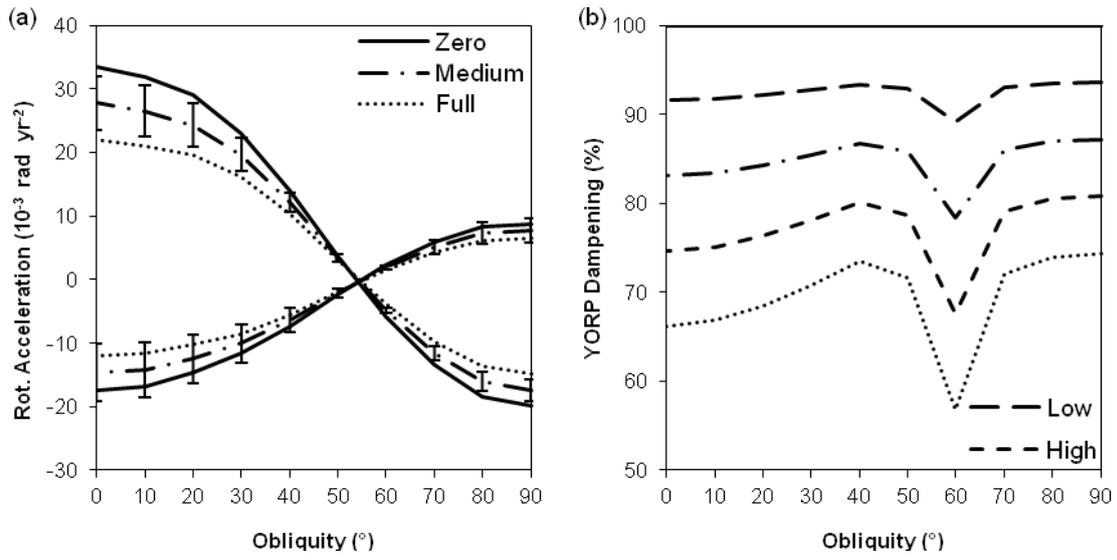

Figure 11:

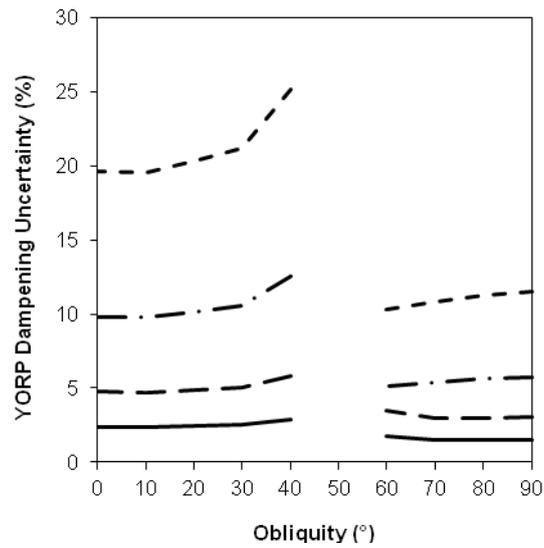

Figure 12:

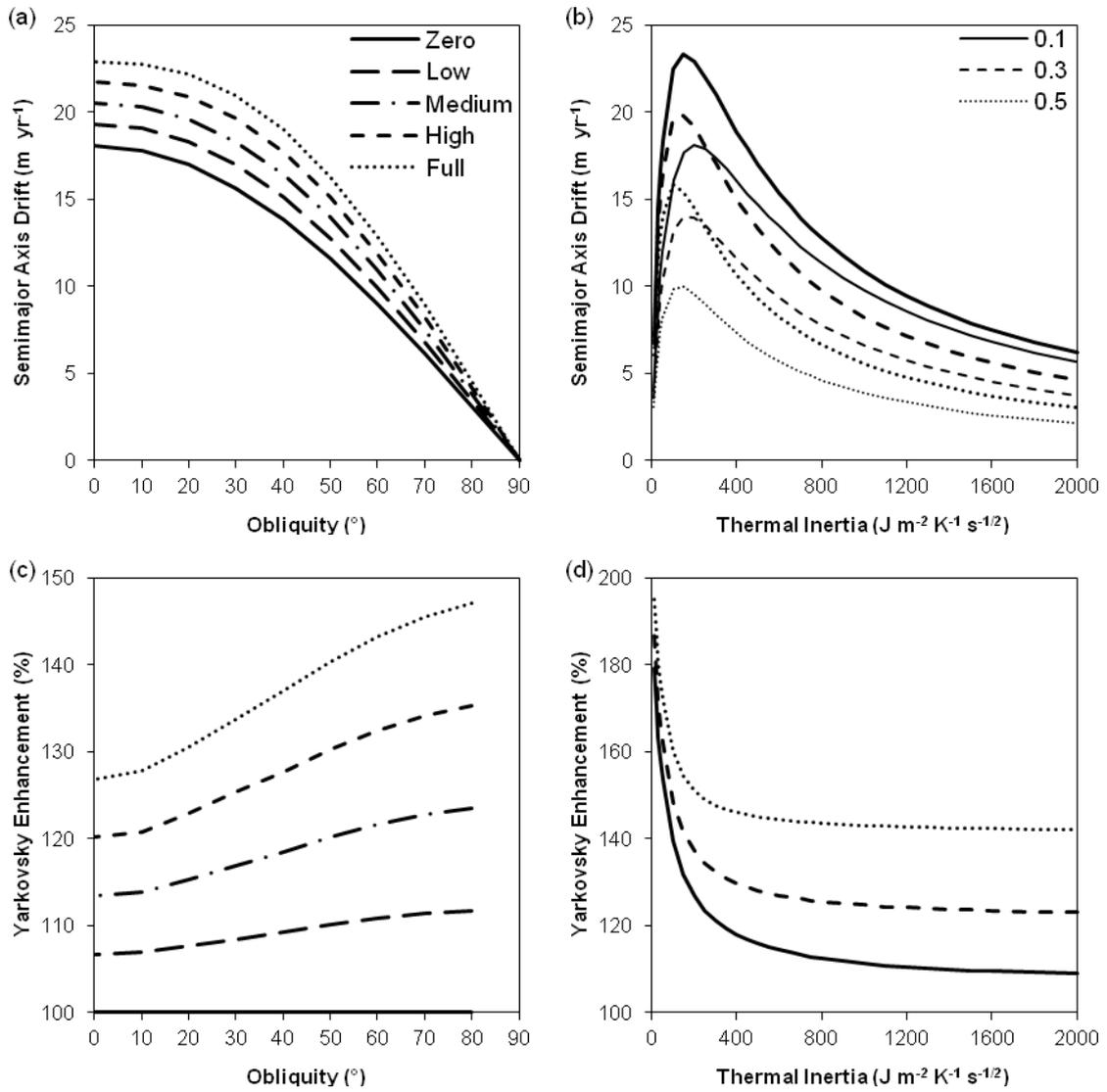

Figure 13:

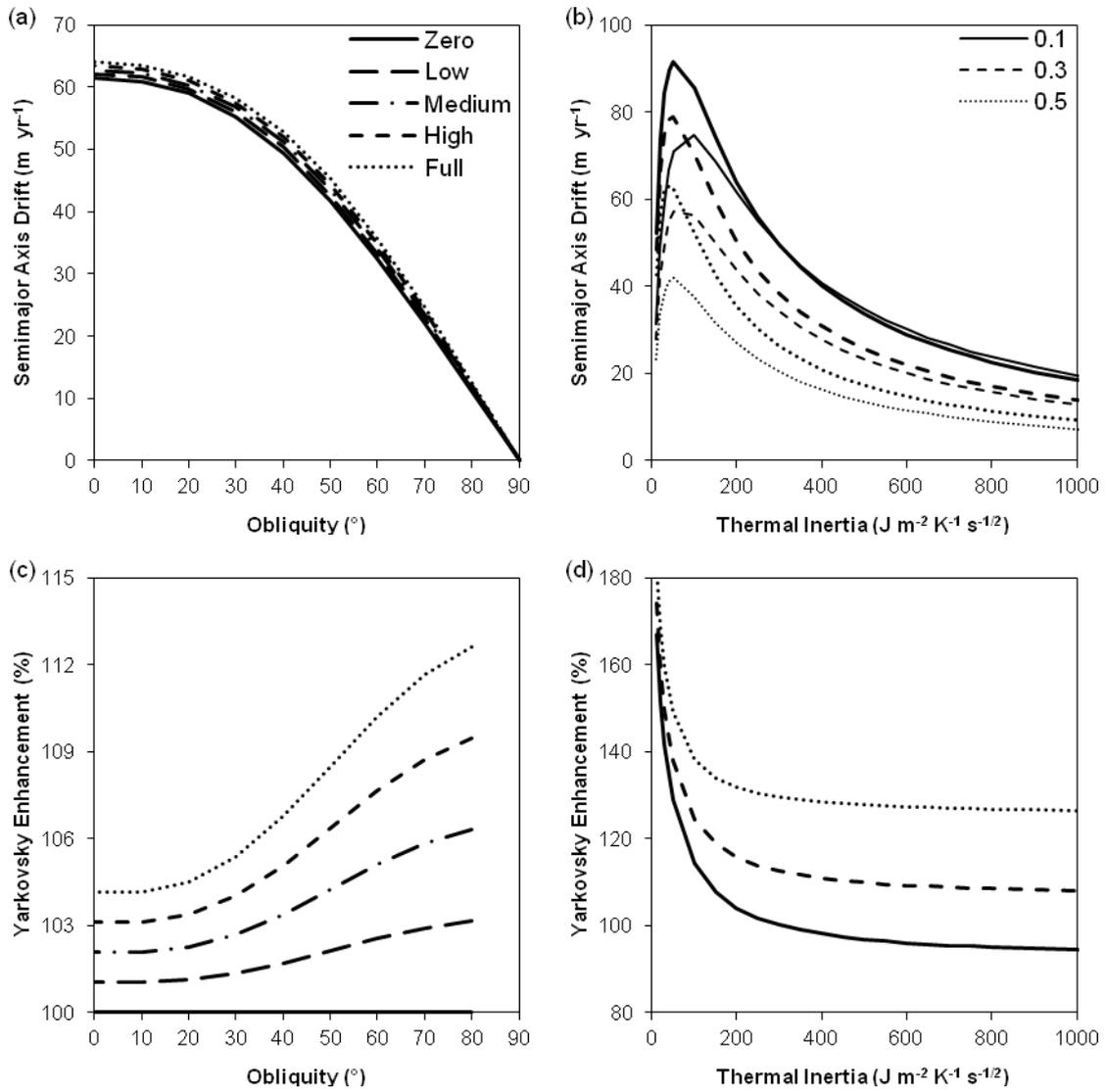

Figure 14:

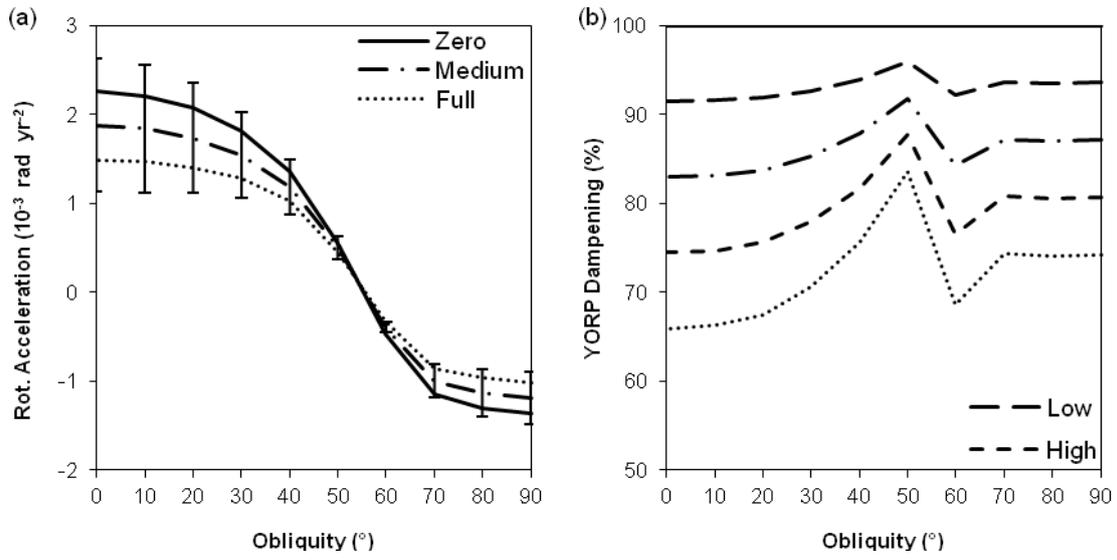

Figure 15:

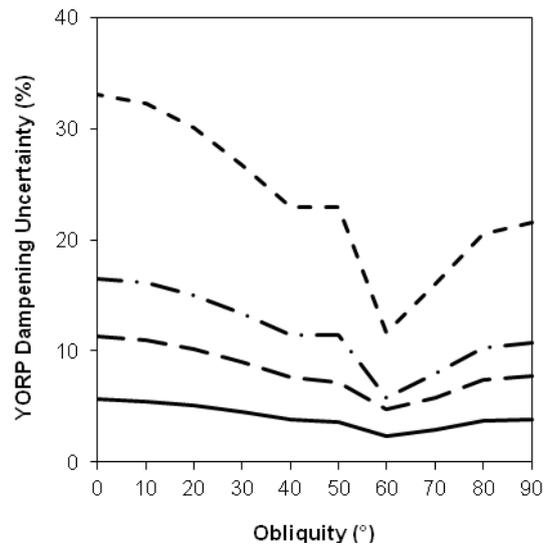

Figure 16:

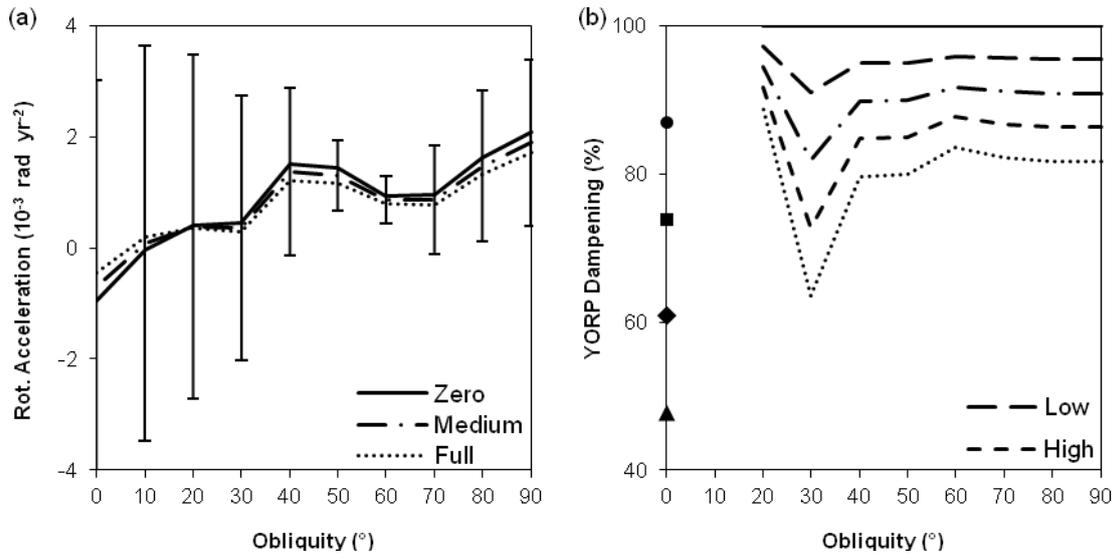

Figure 17:

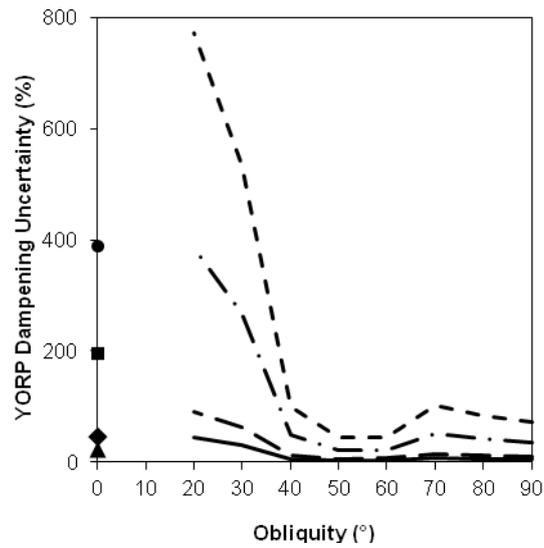

Figure 18:

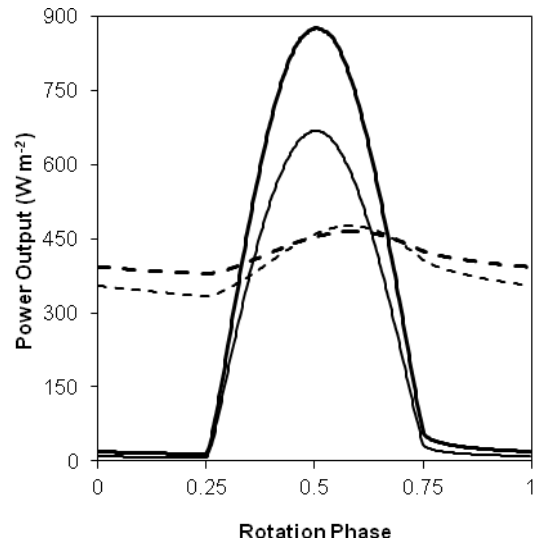

Figure 19:

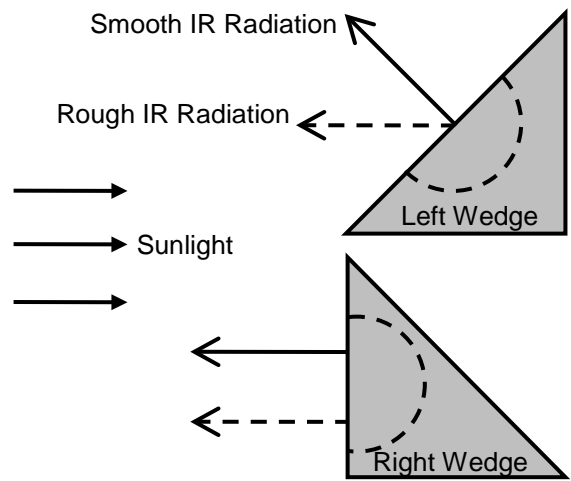

Figure 20:

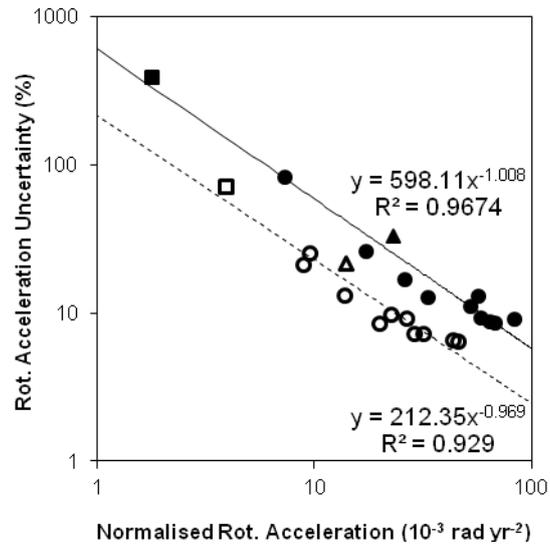

Figure B1:

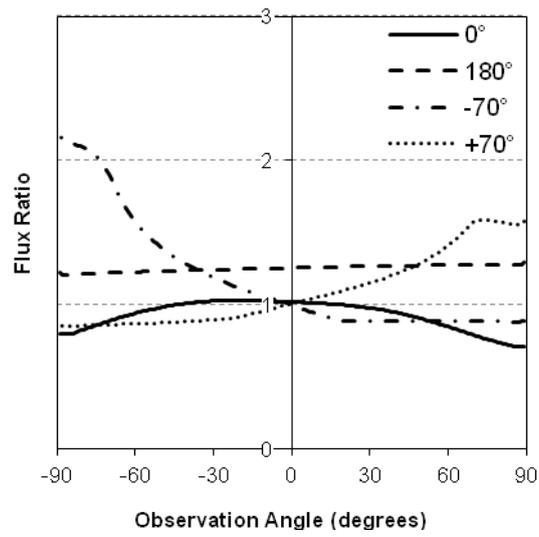